\documentclass[useAMS,usenatbib]{mn2e}

\usepackage{rotating}
\usepackage{graphicx}
\usepackage{txfonts}
\usepackage{lscape}
\usepackage{natbib}
\bibpunct{(}{)}{;}{a}{}{,} % to follow the A&A style

\title[The Physics and Mass Assembly of distant galaxies with the
  E-ELT]{Simulating the physics and mass assembly of distant galaxies
  out to z$\sim$6 with the E-ELT} \author[M. Puech et al.]{M.
  Puech$^{1,2}$\thanks{E-mail: mathieu.puech@obspm.fr}, P. Rosati$^{1}$, S.
  Toft$^{1,3}$, A. Cimatti$^4$, B. Neichel$^{5,2,6}$, \& T.
  Fusco$^{5}$\\ $^{1}$ESO, Karl-Schwarzschild-Strasse 2, D-85748
  Garching bei M\"unchen, Germany\\ $^{2}$GEPI, Observatoire de Paris,
  CNRS, University Paris Diderot; 5 Place Jules Janssen, 92190 Meudon,
  France\\ $^{3}$Dark cosmology centre, Niels Bohr Institute,
  University of Copenhagen, Juliane Mariesvej 30, DK-2100 Copenhagen,
  Denmark\\ $^{4}$ Dipartimento di Astronomia, Alma Mater Studorum,
  Universita di Bologna, Via Ranzani 1, 40127, Italy\\ $^{5}$ONERA, BP
  72, 92322 Chatillon Cedex, France\\ $^{6}$Gemini Observatory, Colina
  El Pino s/n, Castilla 603, La Serena, Chile\\}
\begin{document}

\date{Accepted ...}

\pagerange{\pageref{firstpage}--\pageref{lastpage}} \pubyear{2002}

\maketitle

\label{firstpage}

\begin{abstract}
One of the main science goal of the future European Extremely Large
Telescope will be to understand the mass assembly process in galaxies
as a function of cosmic time. To this aim, a multi-object, AO-assisted
integral field spectrograph will be required to map the physical and
chemical properties of very distant galaxies. In this paper, we
examine the ability of such an instrument to obtain spatially resolved
spectroscopy of a large sample of massive ($0.1 \leq M_{stellar} \leq
5\times10^{11}M_{\odot}$) galaxies at $2\leq z < 6$, selected from
future large area optical-near IR surveys. We produced a set of about
one thousand numerical simulations of 3D observations using reasonable
assumptions about the site, telescope, and instrument, and about the
physics of distant galaxies. These data-cubes were analysed as real
data to produce realistic kinematic measurements of very distant
galaxies. We then studied how sensible the scientific goals are to the
observational (i.e., site-, telescope-, and instrument-related) and
physical (i.e., galaxy-related) parameters. We specifically
investigated the impact of AO performance on the science goal. We did
not identify any breaking points with respect to the parameters (e.g.,
the telescope diameter), with the exception of the telescope thermal
background, which strongly limits the performance in the highest
(z$>$5) redshift bin. We find that a survey of $N_{gal}$ galaxies that
fulfil the range of science goals can be achieved with a $\sim$90
nights program on the E-ELT, provided a multiplex capability $M \sim
N_{gal}/8$.
\end{abstract}


\section{Introduction}
Over the last decade, the synergy of 8-10 meter class telescopes with
HST has strongly re-invigorated the field of galaxy formation and
evolution by unveiling very distant galaxies up to z$\sim$6 (e.g.,
\citealt{cuby03,rhoads03,bremer04,bouwens06}), by allowing the first
determination of the global star formation history since redshift
z$\sim$6 (e.g., \citealt{hopkins04}), and by providing the first
insights on the stellar mass assembly history out to z$\sim$5 (e.g.,
\citealt{drory05,pozzetti07,marchesini07,perez07}). Despite these
recent progresses, the outstanding question remains on how and when
galaxies assembled their baryonic mass across cosmic time. The CDM
standard model has provided a satisfactory scenario describing the
hierarchical assembly of dark matter halos, in a bottom-up sequence
which is now well-established over the whole mass structure spectrum.
In contrast, little progress has been made in the physical
understanding of the formation and evolution of the baryonic component
because the conversion of baryons into stars is a complex, poorly
understood process.

As a result, all intellectual advances in galaxy formation and
evolution over the last decade have been essentially empirical, often
based on phenomenological (or semi-analytical) models, which heavily
rely on observations to describe, with simplistic rules, such
processes as star formation efficiency, energy feedback from star
formation and AGN, chemical evolution, angular momentum transfer in
merging, etc. Cornerstones observations in this empirical framework
are the total and stellar mass of galaxies and their physical
properties, including the age and metallicities of their underlying
stellar populations, dust extinction, star formation rate, and
structural/morphological parameters. The study of well-established
scaling relations involving a number of these physical parameters
(e.g. mass-metallicity, fundamental plane, colour-magnitude,
morphology-density) are essential for understanding the physical
processes driving galaxy evolution. However, with the current
generation of 10m-class telescopes, we have been able to construct for
example the fundamental plane of early-type galaxies, or to measure
the Tully-Fisher relation of late-type galaxies over a wide range of
masses only at low and intermediate redshifts (z$<$1), whereas only
the brightest or most massive galaxies have been accessible at z $>$
1, and a direct measurement of masses has been almost completely out
of reach at z $>$ 2. Thus, our ability to explore the evolution and
origin of the aforementioned scaling relations has rapidly reached the
limit of 10m-class telescopes.

Hence, most of the outstanding questions arisen from recent
observational galaxy evolution studies which have pushed the 10m-class
telescopes to their limits call for an Extremely Large Telescope
(ELT), specifically to extend the spectroscopic limit by at least two
magnitudes with near-diffraction limit angular resolution. IFU
spectrographs on 8-10m class telescopes (e.g., VLT/SINFONI and
Keck/Osiris) are now routinely deriving the spatially-resolved
kinematics of massive (i.e., with stellar masses larger than
10$^{10}M_\odot$), distant galaxies, from z$\sim$0.4 to z$\sim$3
\citep{flores06,puech06,yang08,forster06,wright07,shapiro08,bournaud08,genzel08,vanstarkenburg08,law09,wright09,epinat09,forsterschreiber09,lemoine09}.
Such studies have brought new and essential insights into galaxy
evolution processes, such as the fraction of rotators as a function of
redshift \citep{yang08,forsterschreiber09}, a better understanding of
the evolution of the Tully-Fisher relation
\citep{puech08,cresci09,puech09}, or the first glimpse into the
evolution of the angular momentum \citep{puech07,bouche07}. However,
several uncertainties and limitations remain, especially at the
highest redshifts (i.e., z$>$1). For instance, z$\sim$2 surveys were
drawn from different selection criteria (e.g., BX, BM, or BzK
galaxies), and this might bias the resulting sample in a subtle and
quite uncontrolled way. Besides, samples in the NIR are selected in
optimal atmospheric windows, free of OH sky lines and maximising the
atmospheric throughput, which intrinsically limits the redshift range
and size of the resulting sample. The only way to overcome these
issues is to move to telescopes with larger collecting areas.

ESO is currently developing a Phase B study for a European 42-meter
telescope \citep{gilmozzi08,spyromilio08}. As part of the project
development, the Design Reference Mission (DRM) aims at producing a
set of observing proposals and corresponding simulated data which
together provide the project with a tool to assess the extent to which
the E-ELT addresses key scientific questions and assist in critical
trade-off decisions\footnote{see
  http://www.eso.org/sci/facilities/eelt/science/drm/}. In the frame
of the DRM effort, simulations of 3D high-z galaxy observations were
undertaken. These observations are expected to yield direct kinematics
of stars and gas in the first generation of massive galaxies (in the
range 0.1 $\leq$ M $\leq$ 5$\times$10$^{11}$M$_\odot$), as well as
their stellar population properties. This will allow one to derive
dynamical masses, ages, metallicities, star-formation rates, dust
extinction maps, to investigate the presence of disk and spheroidal
components and the importance of dynamical processes (e.g. merging,
in/outflows) which govern galaxy evolution. These data will also allow
one to study the onset of well known scaling relations at lower
redshifts, and to witness the gradual shift of star formation from the
most massive galaxies in the highest density regions to less massive
galaxies in the field.

To assess the science achievement of an NIR IFS on the E-ELT, as well
as to better understand the interactions between the telescope, site,
and instrument, we have produced an extensive set of $\sim$1000
simulations of observations of very distant galaxies. It is important
to realize that if one wants to study what kind of instrument is
needed to understand the galaxy mass assembly process, then one needs
to explore a huge parameter space, i.e., from relaxed smooth or clumpy
rotating disks to major and minor mergers, taking care of spanning all
possible mass ratios, viewing angles, merger geometry and so on, which
would make such a direct approach very difficult, and probably even
impossible, on the practical side. Moreover, one has to recognise that
our knowledge and understanding of the physics of distant galaxies is
still incomplete, which necessitates to extrapolate some of their
characteristics. Therefore, the simulations presented in this paper do
not intend to be exhaustive in any sense. Rather, we aim at exploring
this very large parameter space using reasonable assumptions and
guesses. Indeed, the DRM exercise consisted in a broad exploration of
the parameter space in realistic observing conditions, in order to
identify possible limitations and/or breaking points, as well as
interactions between the telescope, site, and instrument, which could
potentially impact the telescope design.

This paper is the second of a series that explores the performances of
a NIR IFS on the E-ELT. In the first paper, we presented our
methodology to simulate realistic observations of distant galaxies
\citep{puech08b}. We also produced first simulations and explored
performance using a few scientifically-motivated cases. They
illustrated the concept of ``scale-coupling'', i.e., the relationship
between the IFU pixel scale and the size of the kinematic features
that need to be recovered by 3D spectroscopy in order to understand
the nature of the galaxy and its substructure. In \cite{puech08b}, we
focused on the largest spatial scales, which are of particular
interest because they carry most of the kinematic information useful
to reveal the process underlying galaxy dynamics, i.e., whether a
given galaxy is in a coherent and stable dynamical state (e.g.,
rotation), or, on the contrary, out of equilibrium (e.g., subsequently
to a merger).

In this paper, we present an updated version of the simulation
pipeline. The main improvement is related to the inclusion of thermal
emissions from both the IFS and the telescope. This allowed us to
explore a wider parameter space, and especially areas where
observations are limited by the thermal emission rather than by the
sky background (e.g., relatively faint targets in the K band). We also
significantly extended the range of morpho-kinematic templates and of
physical parameters (e.g., radius, mass, and velocity as a function of
redshift) considered in the simulations. This paper is organised as
follows. In Sect. 2, we present our methodology and the pipeline used
for simulations. In Sect. 3, we detail the scientific and
observational inputs used in the simulations. In Sect. 4, we present
the results of the simulations, which are discussed in Sect. 5. A
conclusion is drawn in Sect 6. Throughout, we adopt $H_0=70$ km/s/Mpc,
$\Omega _M=0.3$, and $\Omega _\Lambda=0.7$, and the $AB$ magnitude
system.

\section{The physics and mass assembly of galaxies out to z$\sim$6 with the E-ELT}
%This entire section should be more or less copied from the proposal.

\subsection{Goals of the simulations}
%List here the main questions to be answered by the simulations.
Simulations presented in this paper focus on the sub-sample of
distant emission line galaxies: due to Signal-to-Noise Ratio (SNR)
limitations, absorption line galaxies at z $\gtrsim$ 1.5 will more
likely be studied using integrated spectroscopy (see DRM Science Case
C10-3 ``ELT integrated spectroscopy of early-type galaxies at z $>$
1''\footnote{http://www.eso.org/sci/facilities/eelt/science/drm/C10/}). Therefore,
in the following, only emission line galaxies will be considered and
for convenience, we will sometimes use the single word ``galaxies'' to
refer to ``emission line galaxies''.

Spatially-resolved kinematics of such galaxies provides a useful
test-bed for 3D spectrographs, because it drives the stringent
requirements on the SNR: while flux is the zero-order moment of an
emission line, the velocity and velocity dispersion are derived from
the position and the width of emission lines, which are their first
and second moments, and higher order moments are always characterised
by larger statistical uncertainties. Hence, simulations shall assess
the principal scientific goal of spatially-resolved spectroscopy of
distant emission line galaxies, focusing on kinematics. Different
objectives can be defined depending on the level of accuracy and/or
spatial scale one wants to probe; we defined the following objectives
(dubbed here as ``steps''), ordered by increasing
complexity/difficulty:
\begin{enumerate}
\item Simple (3D) detection of emission line galaxies: what stellar
  mass can be reached at a given SNR as a function of redshift, AO
  system, environmental conditions, etc.?

\item Recovery of large scale motions (see, e.g., \citealt{flores06}):
what are the conditions under which it is possible to recover the
dynamical state of distant galaxies (e.g., relaxed rotating disks vs.
non-relaxed major mergers)?

\item Recovery of Rotation Curves (RC): what are the conditions under
  which it is possible to recover the rotation velocity V$_{rot}$ (to
  derive, dynamical masses or the Tully-Fisher relation, see, e.g.,
  \citealt{puech06,puech08}) or the whole shape of the RC (for
  derivation and decomposition in mass profiles, see, e.g.,
  \citealt{blaisouelette01})?

\item Recovery of the detailed kinematics: what are the conditions
under which it is possible to detect internal structures in distant
galaxies, like clumps in distant disks (see, e.g.,
\citealt{bournaud08})?
\end{enumerate}

Studying and evaluating in detail the future achievement and
measurement accuracy of an NIR integral-field spectrograph on the
E-ELT in these areas is beyond the scope of this paper. Moreover, in
addition to the scientific limitations pointed out in the
Introduction, one has to take into account the fact that technical
developments (both on the telescope and instrument side) are not
advanced enough to allow us to conduct such a detailed study, since
several key elements are currently not fully known (e.g., the thermal
contribution from the telescope, coatings, instrument design). Rather,
the goal of this paper is to restrict the acceptable underlying
parameter space (see below), and identify breaking points on the
technical side that might strongly impact the achievable science of
an IFS an the E-ELT.

However, to put this study on as realistic as possible grounds, an
observational proposal was defined prior to simulations. This program
was designed to provide us with an ultimate test of galaxy formation
theories. It is therefore ambitious, with the goal of obtaining
spatially-resolved kinematics of about one thousand of galaxies
spanning a large range of cosmic time and mass. Current 3D surveys on
8-10m telescopes already allowed us to sample a significant part of
the cosmic look-back time, from z=0.4 to z$\sim$3, i.e., up to
$\sim$11 Gyr ago. Main 3D samples are concentrated around z$\sim$0.6
(see the IMAGES sample, \citealt{yang08}) and z$\sim$2 (see the SINS
and OSIRIS samples, \citealt{forsterschreiber09,law09}), and it is
unlikely that current or event future 3D spectrographs on 8-10m
telescopes will provide us with large and representative samples
beyond z$\sim$3, due to limitations in surface brightness detection,
even when fed with adaptive optics. Therefore, we focused on the
redshift range 2-5.6, which remains largely unexplored, with the
advantage of sharing the z$\sim$2 limit with existing data. The upper
limit is driven by the availability of emission lines in the NIR
window, and corresponds to the [OII] emission line getting out the K
band.

It is worth emphasising that such a 3D survey is by nature radically
different from what is currently done at z$>$1. Because of the limited
sensitivity of current 3D spectrographs, galaxies are targeted in
optimal windows, where the atmospheric absorption is low. This,
combined to the collection of selection criteria used at high-z to
pre-select targets with known redshifts (e.g., BzK, or Lyman-break
techniques) result in samples whose selection functions and
representativity have been the subject of a long debate. The only way
to overcome these limitations is to conduct a deep optical-NIR imaging
and redshift survey with high completeness up to z$\sim$5.6, from
which a 3D follow-up on the E-ELT will guarantee, \emph{without any
  pre-selection}, the representativity of the resulting sample.

Several integral field spectrographs (IFS) are currently under study
for the E-ELT. The present program is designed for a multi-object IFS,
since multiplex is required to reach a significant number of targets.
For example, EAGLE, the project of multi-integral field spectrograph
for the E-ELT, would be a well-suited instrument for such a survey
\citep{cuby08,puech09b}. However, we want to emphasise that the
present paper does not intend to explore the scientific capabilities
of this particular instrument, beyond the fact that it is a NIR
multi-IFU spectrograph.

\subsection{Simulation pipeline}
The end-products of 3D spectrographs are usually data-cubes in FITS
format. Hence, we have developed a simulation pipeline that produces
such mock data. These simulated data are produced assuming a perfect
data reduction process, whose impact has therefore been neglected (see
Sect. 5.1).

Data-cubes are simulated using a backward approach of the usual
analysis of 3D data. During such an analysis, each spaxel is analysed
separately in order to extract from each spectrum the continuum and
the first moments of emission lines. One ends up with a set of maps
describing the spatial distribution of the continuum and line
emissions (zeroth order moment of the emission line), as well as the
gas velocity field (first order moment) and velocity dispersion map
(second order moment). Such maps are routinely derived from
Fabry-Perot observations of local galaxies (see, e.g,
\citealt{epinat08}), or can be generated as by-products of
hydro-dynamical simulations of local galaxies (e.g., \citealt{cox06}).
Using simple rules and empirical relations, these maps can be rescaled
(in terms of size and total flux) to provide a realistic description
of distant galaxies (see Sect. 3.1). Assuming a Gaussian shape for all
spectra, it is then straightforward to reconstruct a data-cube from
such a set of four maps, by ``reversing'' the usual analysis of 3D
data. This allows us to produce a data-cube at high spatial
resolution, which can then be degraded at the spatial resolution (AO
PSF) and sampling (IFU pixel scale) required for simulating data
produced by a given 3D spectrograph. During the process, realistic sky
and thermal backgrounds, as well as photon and detector noise are
added. The pipeline is fully described in \cite{puech08b} and
\cite{puech08c}.

The simulation pipeline generates $ndit$ data-cubes with individual
exposure time of $dit$, which are combined by estimating the median of
each pixel to simulate several individual realistic exposures. Since
we have only included random noise, it is similar to having dithered
all of individual exposures and combining them after aligning them
spatially and spectrally. Sky frames are evaluated separately (i.e.,
with a different noise realization), and then subtracted from each
individual science frame, reproducing the usual procedure in both
optical and NIR spectroscopy.

\subsection{Overview of simulation outputs}
We list here the FITS files generated by the simulation pipeline that
are relevant for the present study:
\begin{itemize}
\item An IFU data-cube: this is the final product of the simulation,
which corresponds to mock observations;

\item A total background spectrum (sky continuum, OH sky lines, and
total thermal background);

\item A thermal background spectrum;

\item An SNR data-cube, which gives the expected SNR within each pixel
  of the IFU data-cube. This spectroscopic SNR is derived as follows:
$$
SNR(i_x,j_y,k_{\lambda})=\frac{O(i_x,j_y,k_{\lambda})*\sqrt{ndit}}{\sqrt{O(i_x,j_y,k_{\lambda})+S(i_x,j_y,k_{\lambda})+ron^2+dark}},
$$ where $O(i_x,j_y,k_{\lambda})$ and $S(i_x,j_y,\lambda)$ are
respectively the object and sky flux per $dit$ (after accounting for
atmospheric transmission) in the spatial position $(i_x,j_y)$ of the
data-cube (in pixels), and at the spectral position $k_{\lambda}$
along the wavelength axis (in pixels). In the following, the ``maximal
SNR in the emission line in the pixel $(i_x,j_y)$'' refers to
MAX$_{k_{\lambda}}$[$SNR(i_x,j_y,k_{\lambda})$], and the
``spatial-mean SNR'' refers to the average of this quantity over the
intrinsic galaxy diameter, i.e., the diameter of the galaxy
irrespective of what parts of the galaxy are detected.
\end{itemize}
For each simulation, the main parameters are recorded in the headers
of the corresponding FITS files.\footnote{Examples of simulated
  data-cubes can be downloaded at
  http://www.eso.org/sci/facilities/eelt/science/drm/C10/.}

The simulated IFU data-cubes are then analysed using an automatic data
analysis pipeline similar to those generally used to analyse data of
high redshift galaxies. During this process, each spatial pixel of the
simulated data-cube is fitted with a Gaussian in wavelength, whose
position and width correspond respectively to the velocity and
velocity dispersion of the gas in this spatial pixel. Only pixels with
a kinematic signal to noise ratio $SNR_{kin}$ of at least three (see
definition below) were considered to limit uncertainties. A more
detailed description of the process can be found in \cite{puech08b}
and \cite{puech08c}. The analysis pipeline produces the following FITS
files:
\begin{itemize}
\item An emission line flux map;

\item A velocity field;

\item A velocity dispersion map;

\item A map of the kinematic $SNR_{kin}$, which is defined as the
  total flux in the emission line divided by the noise on the
  continuum $\sigma _{continuum}$ and $\sqrt N_{pix}$, the number of
  pixels within the emission line;

\item A map of the maximal $SNR$, i.e., the $SNR$ at the peak of the
emission line within each spaxel of the simulated IFU data-cube. A
spatial-mean value is written down into the header (see above).
\end{itemize}

\subsection{Methodology for simulations}

All input parameters can be separated into two broad categories,
namely the ``physical'' parameter space, which includes all parameters
defining the distant galaxy (i.e., redshift $z$, galaxy diameter,
continuum AB magnitude $m_{AB}$, rest-frame emission line equivalent
width $EW_0$, velocity gradient, and morpho-kinematic type), and the
``observational'' parameter space, which includes the telescope
(primary mirror M1 and secondary mirror M2 diameters, temperature
$T_{tel}$, and emissivity $\epsilon _{tel}$), the instrument (spectral
resolution $R$, IFU pixel size $\Delta_{pix}$, detector integration
time $dit$, number of exposures $ndit$, temperatures $T_{instr}$ and
emissivities $\epsilon _{instr}$, AO correction, and global
transmission t$_{transm}$), and the site (seeing, $C_n^2$ profile,
outer scale of the turbulence $L_0$, sky brightness and atmospheric
transmission).

Given the very large number of parameters to be investigated, as well
as the very large range of values to be explored, it is useful to
define a ``reference case'', around which the parameter space can be
explored and compared with. As such a reference case, we adopted an
M$_*$ galaxy at z=4 observed using MOAO (Multi-Object Adaptive Optics,
see Sect. 3.2) with a median seeing of 0.8 arcsec. At this redshift,
the [OII] emission line is observed in the H-band, where the influence
of the thermal background is minimised in comparison with the K-band.
This makes this reference case as independent as possible of the
telescope design (e.g., number of mirrors), environmental conditions
(site selection), and instrument characteristics (e.g., number of warm
mirrors), which are not all fully known at present.

This reference case will be used to assess separately the influence of:
\begin{itemize}
\item the AO correction for a given set of other
observational and physical parameters;

\item the observational parameters for a given set of
physical parameters and AO correction;

\item the physical parameters for a given set of
observational parameters (AO correction included).
\end{itemize}

\subsection{Metrics and figures of merit}
To decide whether or not the scientific goals have been met, we
adopted a pragmatic point of view and defined as a general metric the
total observation time of the survey $T_{intg}$ required to achieve
the proposal goals, i.e., observe $N_{gal}$ galaxies more massive than
$M_{stellar}=10^{10} M_\odot$ in the redshift range 2-5.6. A high
number of galaxies ($N_{gal} \gtrsim$100) is required to allows us to
derive statistics over the morpho-kinematic types of galaxies in
several redshift and stellar mass bins (see Sect. 4.6). We require
$T_{intg} \leq 100$ nights, which roughly corresponds to the total
time allocated at ESO to Large Programs per VLT per year (i.e., 30\%
of the available time). Therefore, such a survey could reasonably be
implemented as a several years effort. Finally, this total time should
correspond to SNR levels that guarantee to reach at least step (ii)
for most of galaxies in the survey, and step (iv) (see Sect. 2.1) for
a more limited sub-sample (to be defined by simulations).

From a more practical point of view, such a survey will be faced with
the difficulty that future targets will have to be drawn from a larger
parent sample because of selection constraints on, e.g., avoidance of
OH sky lines, which reduces the availability of redshift windows at a
30-40\% level in H-band (see \citealt{puech08b}). Other technical
constrains might be related to, e.g., the design of instrument setups
in terms of spectral bandwidth. However, the current development of
EAGLE, the project of multi-integral field spectrograph for the E-ELT,
considers the capability of obtaining the equivalent of large-band
filters in terms of spectral bandwidth in a single observational shoot
\citep{cuby08}. Finally, the construction of this parent sample will
require substantial observing time from current or future facilities
(e.g., VLT/HAWK-I, JWST/NIRCAM, VISTA). Given that the target density
at z$\sim$4 is of the order of one per arcmin$^2$ (down to
$I_{AB}$=25, see \citealt{steidel99,steidel03}, defining a parent
sample of, say, one thousand of galaxies will require deep imaging in
the NIR over several square-degrees. This will be further discussed in
Sect. 5.

\subsection{Test case}

\begin{figure*}
\centering
\includegraphics[width=16cm]{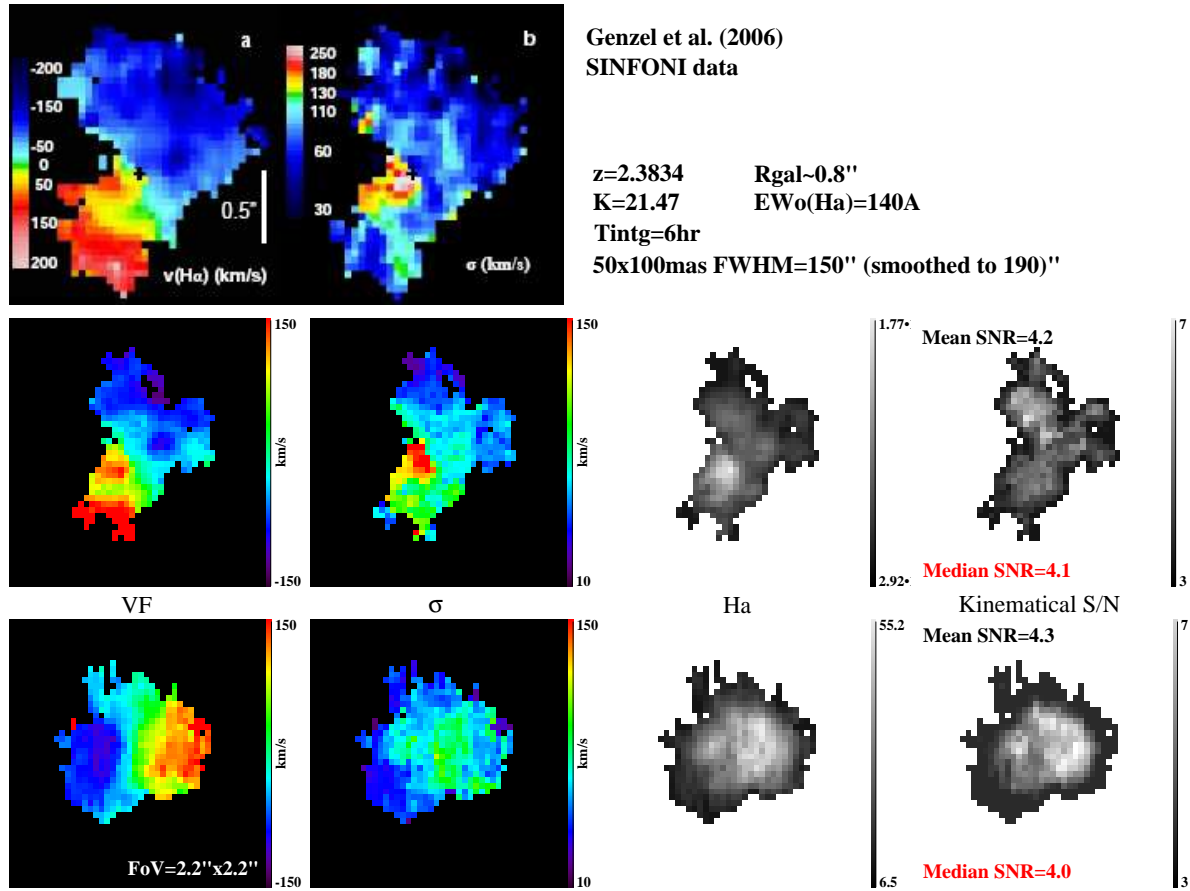}
\caption{Comparison of SINFONI data with simulations of a rotating
  disk (UGC5253 template, see Sect. 3.1) using the same set of
  observational and physical parameters (see text). \emph{First panel:}
  SINFONI observations of a z$\sim$2.4 galaxy (\emph{left:} velocity
  field; \emph{right:} velocity dispersion map), reproduced from
  Genzel et al. (2006). \emph{Second panel:} analysis of the reduced
  SINFONI data-cube (courtesy of N.  F\"orster-Schreiber) using our  own automatic analysis pipeline (see Sect. 2.4). \emph{From left to
    right:} velocity field, velocity dispersion map, H$\alpha$ map,
  and $SNR_{kin}$ map.  \emph{Third panel:} simulations of a rotating
  disk using the observational and physical parameters corresponding to
  the real SINFONI observations. The median $SNR_{kin}$ is found to be
  very similar in both cases.}
\label{sinfoni}
\end{figure*}

We conducted a special first run to compare the results of the
simulation pipeline with real 3D observations on the VLT of a galaxy
at z$\sim$2.4 observed by \cite{genzel06} using SINFONI. The goal of
this run is to use input parameters corresponding to a real observed
case and assess whether or not the simulation pipeline is able to
produce a data-cube with the same quality, quantified using the median
$SNR_{kin}$ (see definition in Sect. 2.3). For the
z$\sim$2.4 galaxy observed with SINFONI, the corresponding input
parameters are: galaxy diameter of 0.8 arcsec, K=21.47,
EW$_0$(H$\alpha$)=140\AA, integration time T$_{intg}$=6 hr, pixel size
of 50$\times$100 mas$^2$, PSF with FWHM=150 milli-arcsec (mas),
temperature of the telescope and instrument of 287K (VLT), emissivity
of the telescope of 6\% (Cassegrain focus), emissivity of SINFONI of
15\%. We took care of mimicking the SINFONI data reduction procedure
by interpolating pixels from the physical 50$\times$100 mas$^2$
spatial scale down to the 50$\times$50 mas$^2$ final scale, and
smoothing the data-cube to a resolution of 190 mas \citep{genzel06} .
The results of this validation run are shown in Fig. \ref{sinfoni} and
demonstrate that, given a complete set of observational and physical
parameters, simulations can produce data-cubes quantitatively similar
to real observations. Moreover, the similarity between the observed
velocity field and velocity dispersion map (see first panel in Fig.
\ref{sinfoni}) and those produced by the automatic analysis pipeline
(see second panel) shows that the automatic analysis pipeline can be
used safely and does not add too much uncertainty on the derived
kinematics compared to a more careful visual examination and fitting
of the data-cube.

\section{Simulations}

\subsection{Scientific inputs}
Scientific inputs are needed in order to (1) provide the simulation
pipeline with morpho-kinematic templates and (2) re-scale these
templates as a function of realistic distant galaxy sizes, fluxes, and
velocity gradients. A global flowchart of the rescaling procedure is
shown in Fig. \ref{rescaling}, while specific details are given below.

\begin{figure}
\centering
\includegraphics[width=8cm]{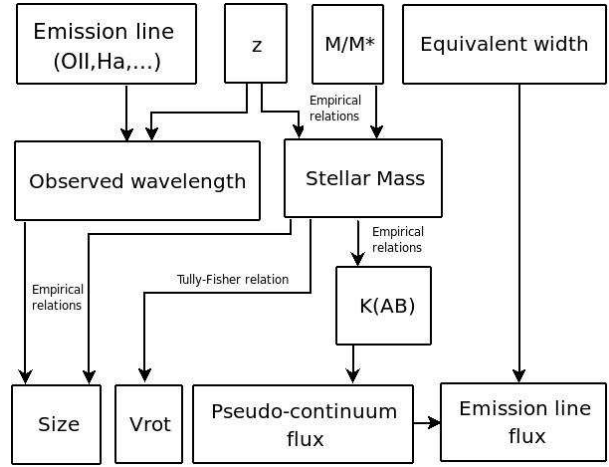}
\caption{Schematic illustration of the method used to re-scale the
morpho-kinematic templates in order to make them match the size and
flux of distant galaxies. Also, the velocity dispersion (not shown
here) can be rescaled using a multiplicative value.}
\label{rescaling}
\end{figure}

\begin{figure*}
\centering
\includegraphics[width=16cm]{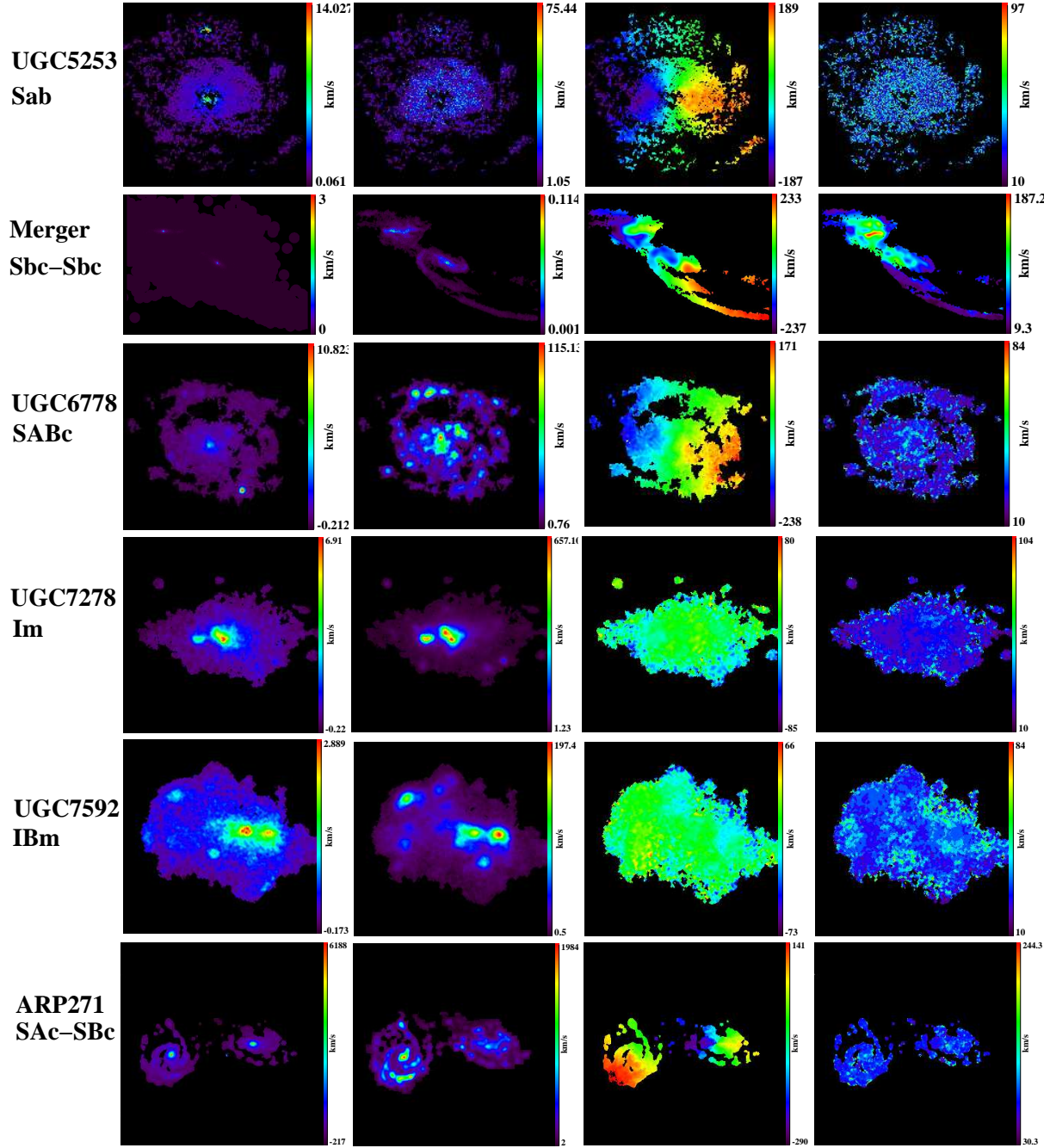}
\caption{Morpho-kinematic templates used for the simulations.
  \emph{From left to right:} continuum map, emission line map, gas
  velocity field, and gas velocity dispersion map. For the major
  merger simulation, the first two maps are the stellar surface
  density and the gas surface density maps (see Tab. \ref{tabtempl}
  for details). Spatial scales are in arbitrary units since the
  templates are rescaled in terms of distant galaxy sizes (see text).}
\label{templates}
\end{figure*}

\begin{figure*}
\centering
\includegraphics[width=16cm]{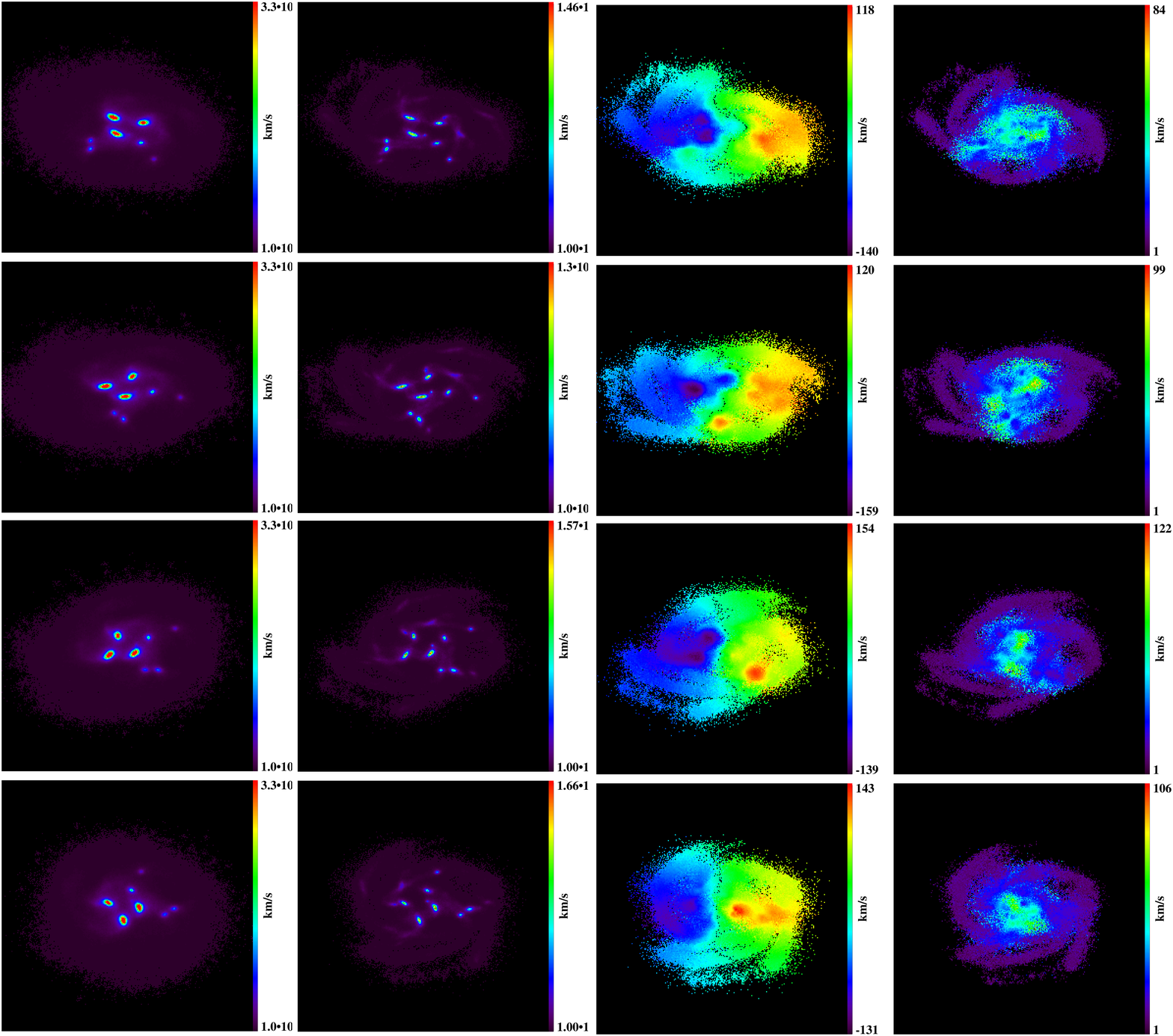}
\caption{Morpho-kinematic templates from simulations of clumpy disks
  viewed from different azimuthal angles (from top to bottom: 0, 45,
  90, and 135 deg). \emph{From left to right:} stellar surface density
  map, gas surface density map, gas velocity field, and gas velocity
  dispersion map (see Tab. \ref{tabtempl} for details). Spatial scales
  are in arbitrary units since the templates are rescaled in terms of
  distant galaxy sizes (see text).}
\label{templates2}
\end{figure*}

{\bf Morpho-kinematic templates:} Table \ref{tabtempl} summarises
the main properties of the high resolution templates used for the
simulations, which are shown in Fig. \ref{templates} and
\ref{templates2}. These templates were obtained from both real
observations (Fabry-Perot interferometry of local galaxies, see
\citealt{epinat08} and \citealt{fuentes04}), and hydro-dynamical
simulations of galaxies (see \citealt{cox06} and
\citealt{bournaud08}).

\begin{table*}
\centering
\begin{tabular}{ccccccc}\hline
Name & Morphological type & $inc$ (deg) & $M_B$ or $M_{stellar}$ & $z$ & Comments & Reference\\\hline
UGC5253 & Sab & 40 & -20.8 & 0.00441 & Rotating disk & \cite{garrido02}\\
UGC6778 & SABc & 30 & -20.6 & 0.003226 & Rotating disk & \cite{garrido02}\\
UGC7278 & Im & 44 & -17.1 & 0.00097 & No rotation & \cite{garrido04}\\
UGC7592 & IBm & 64 & -17.8 & 0.00069 & No rotation & \cite{garrido04}\\
ARP271 & SAc-SBc & 59/32 & -20.6/-21.2 & 0.0087 & Merging pair & \cite{fuentes04}\\
Major merger & Sbc-Sbc & 69 & 5$\times$10$^{10}M_\odot$ & 0.0 & Simulation & \cite{cox06}\\
Clumpy Disks & --- & 50 &  3$\times$10$^{10}M_\odot$ & 1.0 & Simulations & \cite{bournaud08}\\\hline
\end{tabular}
\caption{Main properties of the morpho-kinematic templates used for
  the simulations. The first five templates are real observations,
  while the two last are hydro-dynamical simulations. \emph{From left
    to right:} Name, morphological type, morphological inclination,
  absolute B-band magnitude or stellar mass for simulations, redshift,
  and comments.}
\label{tabtempl}
\end{table*}

{\bf Redshift:} Given the objective of the present study (see
introduction and Sect. 2.1), and as a compromise between the number of
redshift and mass bins at constant total number of targets, three
redshifts were considered for simulations. Table \ref{tabmass}
summarises these redshifts with the corresponding targeted emission
line: we chose z=2 (look-back time of 10 Gyr), z=4 (look-back time of
12 Gyr), and z=5.6 (look-back time of 12.6 Gyr), which samples well
look-back times above z=2. As stated in Sect. 2.3, this choice was not
driven by any consideration of optimal atmospheric transmission, but
purely from scientific grounds. The impact of this selection will be
discussed in Sect. 5.1.2.

For simplicity, the [OII] emission line was assumed to be a single
line centred at 3727\AA; this allows us to avoid introducing an
additional parameter, i.e., the line ratio between the lines of the
doublet, which depends mostly on the electron density in the medium
(e.g., \citealt{puech06}). This does not influence the integrated SNR
over the emission line but leads to overestimate the maximal
spectroscopic SNR in the emission line. The impact of this assumption
will be investigated in Sect. 5.1.1. Note that the highest redshift
considered (z=5.6) corresponds to the limit above which [OII] gets
redshifted out of the K-band.

\begin{figure}
\centering
\includegraphics[width=9cm]{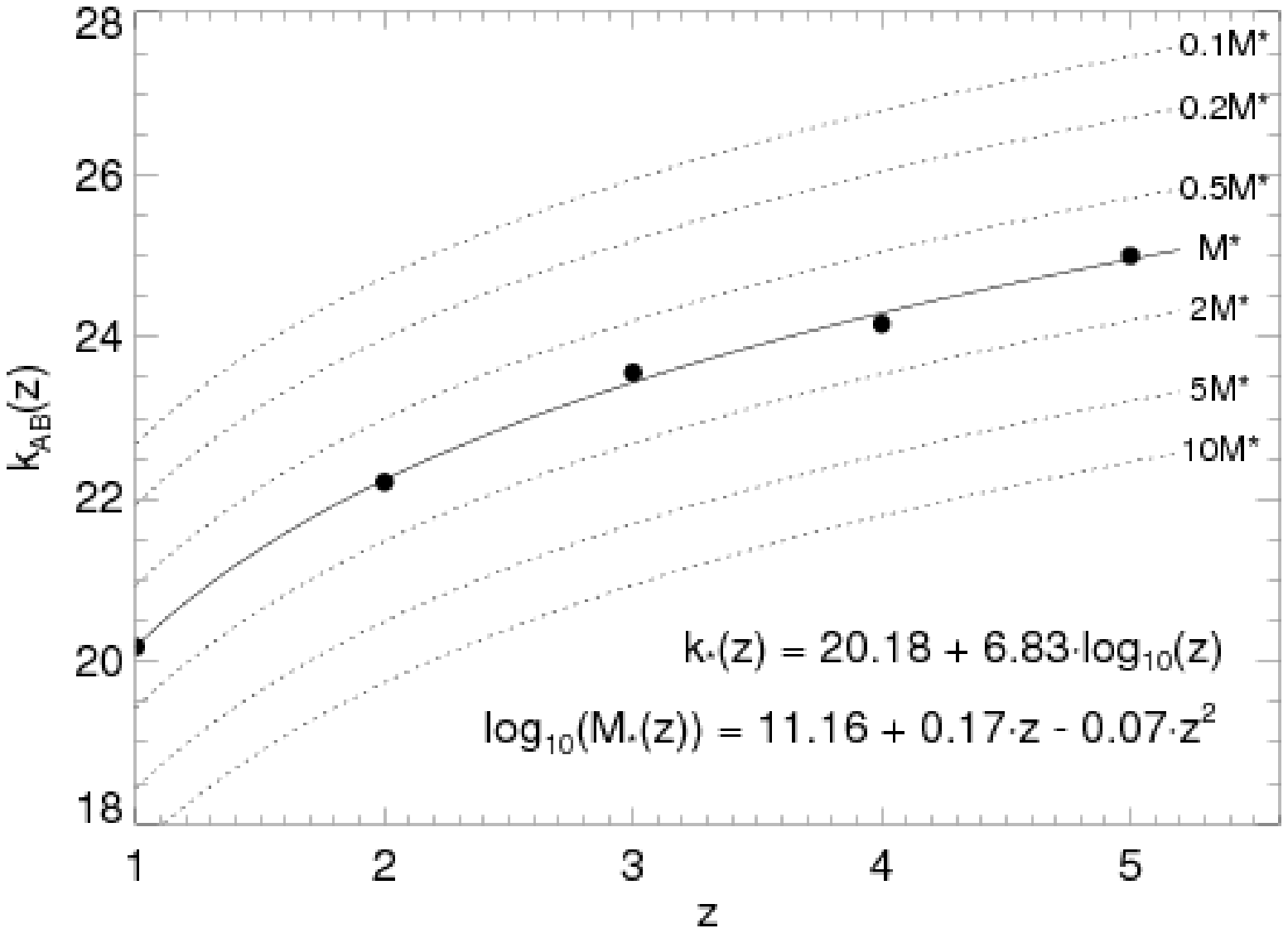}
\caption{Empirical relations between the observed K-band magnitude and
  redshift used to re-scale the morpho-kinematic templates to make
  them match the continuum flux of distant galaxies at the
  corresponding stellar-mass. Black points were derived from fitting
  the luminosity function and mass function of galaxies in the MUSIC
  survey (see text). The variation of M$*$ with redshift, i.e., the
  knee of the GSMF, is given by the formula on the bottom. The upper
  formula gives the empirical evolution of the K-band magnitude of an
  $M_*$ galaxy as a function of redshift. At a given $z$, one finds
  that $K_{AB}=20.18+6.83\log{z}-2.5\log{M_*/M_\odot}$.}
\label{mass}
\end{figure}

{\bf Stellar mass and flux:} We used the MUSIC compilation of public
spectro-photometric data in the GOODS field \citep{grazian06} to
derive empirical relations between redshift, observed K-band
magnitude, and stellar-mass $M_{stellar}$ (see Fig. \ref{mass}). The
latter quantity was expressed as a fraction of the characteristic
stellar-mass $M_*(z)$ at a given redshift, which describes the knee of
the Galaxy Stellar Mass Function (GSMF) at this redshift, according to
a Schechter function. In other words, at a given $z$, galaxies with
stellar mass $M_{stellar}=M_*(z)$ are those which contribute the most
to the stellar mass density at this redshift. Table \ref{tabmass}
gives the corresponding stellar masses in the simulations as a
function of redshift. The pseudo-continuum flux around the emission
line is derived directly from the K-band magnitude. To avoid any
additional parameter, we did not apply any ``colour'' correction
between the rest-frame wavelength of the emission line (e.g., 0.3727
$\mu$m for [OII]) and that of the K-band (e.g., 0.44 $\mu$m at z=4).
Such an assumption is consistent with Spectral Energy Distributions
(SED) of galaxies with a morphological type latter than Sa within a
factor two in flux (see, e.g., \citealt{kinney96}). As a reference
case, we adopted a z=4, $M_*$ galaxy (see Sect. 2.4).

\begin{table*}
\centering
\begin{tabular}{cccccccccc}\hline
z   & Look-back time (Gyr) & Emission line & Observed broad-band & Quantity & 0.1$M_*$ & 0.5$M_*$ & 1.0$M_*$ & 5.0$M_*$ & 10.0$M_*$\\\hline
2.0 & 10 & H$\alpha~\lambda$6563\AA  & K & $log{(M_{stellar}/M_\odot)}$ & 10.2 & 10.9 & 11.2 & 11.9 & 12.2\\
    & & & & $K_{AB}$ & 24.7 & 23.0 & 22.2 & 20.5 & 19.7\\
    & & & & $\Delta V$ (km/s) & 160 & 210 & 260 & 350 & 430\\
    & & & & Size (arcsec)& 0.68 & 1.19 & 1.52 & 2.67 & 3.40\\\hline
4.0 & 12.0 & [OII]$\lambda$3727\AA & H & $log{(M_{stellar}/M_\odot)}$ & 9.7 & 10.4 & 10.7 & 11.4 & 11.7\\
    & & & & $K_{AB}$ & 26.8 & 25.1 & 24.3 & 22.6 & 21.8\\
    & & & & $\Delta V$ (km/s) & 130 & 180 & 200 & 300 & 330\\ 
    & & & & Size (arcsec)& 0.33 & 0.59 & 0.75 & 1.3 & 1.7\\\hline
5.6 & 12.6 & [OII]$\lambda$3727\AA & K & $log{(M_{stellar}/M_\odot)}$ & 8.9 & 9.6 & 9.9 & 10.6 & 10.9\\
    & & & & $K_{AB}$ & 27.8 & 26.0 & 25.3 & 23.5 & 22.8\\
    & & & & $\Delta V$ (km/s) & 90 & 110 & 140 & 200 & 240\\
    & & & & Size (arcsec)& 0.28 & 0.50 & 0.63 & 1.11 & 1.41\\\hline
\end{tabular}
\caption{Redshifts (and look-back times) considered for simulations,
  with the corresponding emission line targeted, and the broad-band
  filter within which the redshifted emission line falls. \emph{On the
    right side:} Stellar mass, K-band magnitude, velocity gradient,
  and size (defined as four times the half-light radius) of the
  simulated galaxies as a function of redshift.}
\label{tabmass}
\end{table*}

{\bf Rest-frame emission line equivalent width:} We assumed
EW$_0$([OII])=30\AA, which is an extrapolation of the mean value found
at z=1 \citep{hammer97}. This parameter does not influence the
emission line flux distribution of the galaxy but is used to set its
total integrated value (see \citealt{puech08b}).

{\bf Size:} We used empirical relations between redshift and
half-light radius from the literature. To mitigate the impact of the
different sample selection criteria used at high redshift, we average
the different values found in the literature
\citep{bouwens04,ferguson04,dahlen07}. The resulting ``mean''
half-light radius was then k-corrected using the empirical relation of
\cite{barden05}, and re-scaled to the assumed stellar-mass using the
local scaling between the K-band luminosity (used as a proxy for
stellar mass) and size reported by \cite{courteau07}, i.e., $R_{half}
\propto L_K^{0.35}$. The total size (diameter) was assumed to be four
times the half-light radius. Note that in the case of a exponential
thin disk of scale-length $R_d$, $R_{25} = 3.2 R_{d}$ and $R_{half} =
1.68 R_{d}$, which approximately leads to $R_{25} \sim 2 R_{half}$.
The diameters (in arcsec) used in the simulations are given in Tab.
\ref{tabmass}.

{\bf Velocity gradient amplitude:}
The velocity amplitude of the velocity field was rescaled using the
local stellar-mass Tully-Fisher relation, following \cite{hammer07}
(see Tab. \ref{tabmass}). Note that strictly speaking, this procedure
is quantitatively correct only if applied to rotating disk
morpho-kinematic templates.

\subsection{Observational inputs}

\begin{figure}
\centering
\includegraphics[width=8cm]{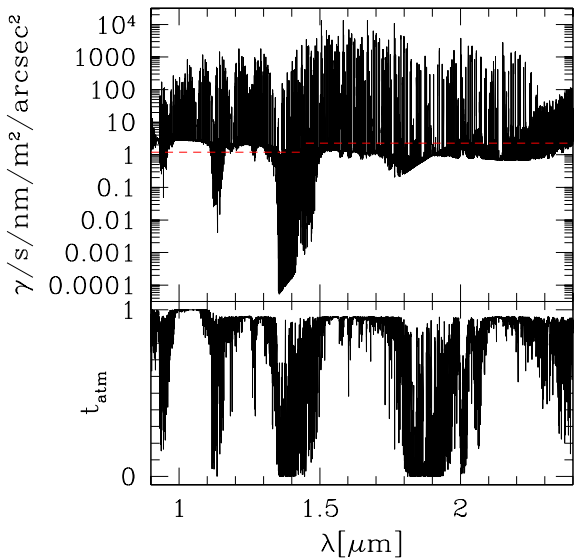}
\caption{Atmospheric model used for the simulations. \emph{Upper
    panel:} sky emission spectrum from the Gemini ETC (see text). The
  dashed-red line is the official DRM model, which is two times
  brighter in H band. \emph{Bottom panel:} Atmospheric model from the
  VLT ETC.}
\label{atm}
\end{figure}

{\bf Site and sky background:} A Paranal-like site is assumed with
T$_{site}$=280K. Atmospheric absorption is modelled following a
Paranal-like site, although at a lower spectral resolution than the
official current model (see bottom panel on Fig. \ref{atm}). Sky
emission (continuum and OH lines) was accounted for using a model from
Mauna Kea, which has the advantage of also including zodiacal
emission, thermal emission from the atmosphere, and an average amount
of moonlight. This Mauna Kea model is two times fainter in H-band than
a Paranal-like site (see upper panel on Fig. \ref{atm}). The influence
of the sky background will be discussed in Sect. 5.2.\footnote{See
  http://www.eso.org/sci/facilities/eelt/science/drm/tech\_data/ for
  a detailed comparison of the two models.}

{\bf Telescope model:} We followed the official 42 meter E-ELT
5-mirror design. Its thermal background was modelled using a gray body,
and assuming an emissivity of 5\% (Ag+Al coating). We neglected the
central aperture, as it is not included in the modelling of the PSFs
(see below). The collecting area of the telescope is therefore 8\%
larger that the official baseline. This will be further discussed in
Sect. 5.1.1.

\begin{figure*}
\centering
\includegraphics[width=7cm]{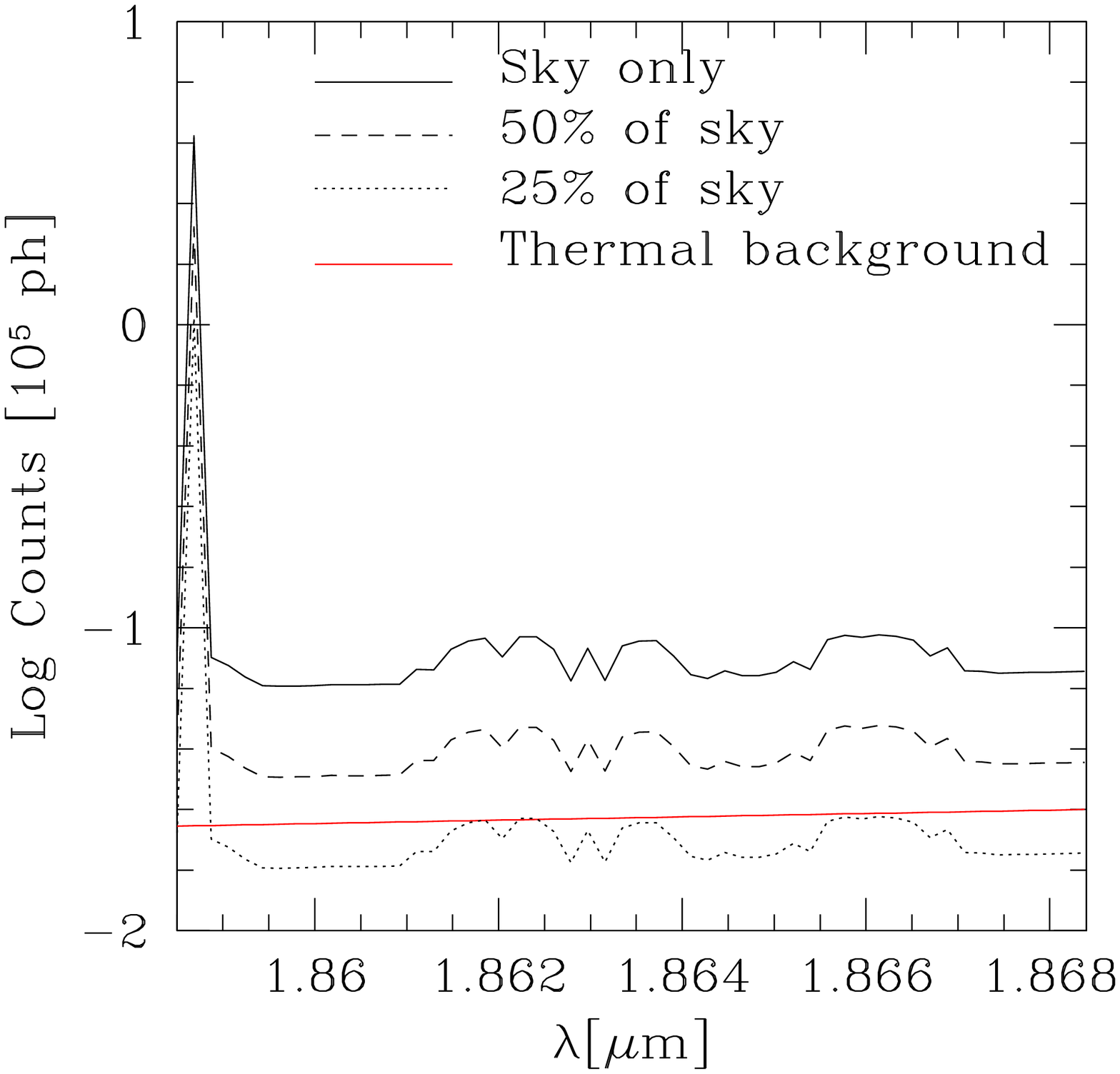}
\includegraphics[width=7cm]{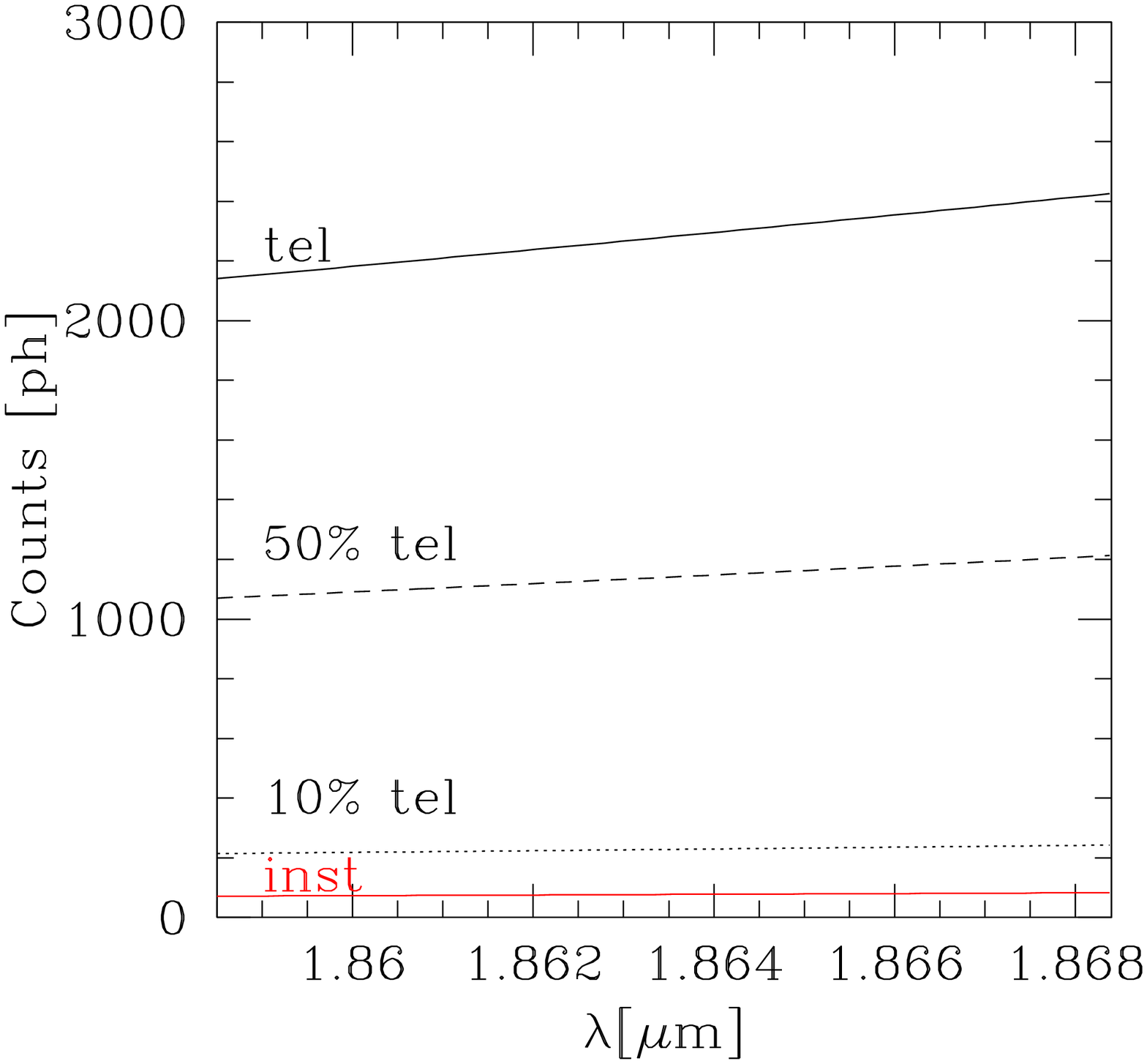}\\
\includegraphics[width=7cm]{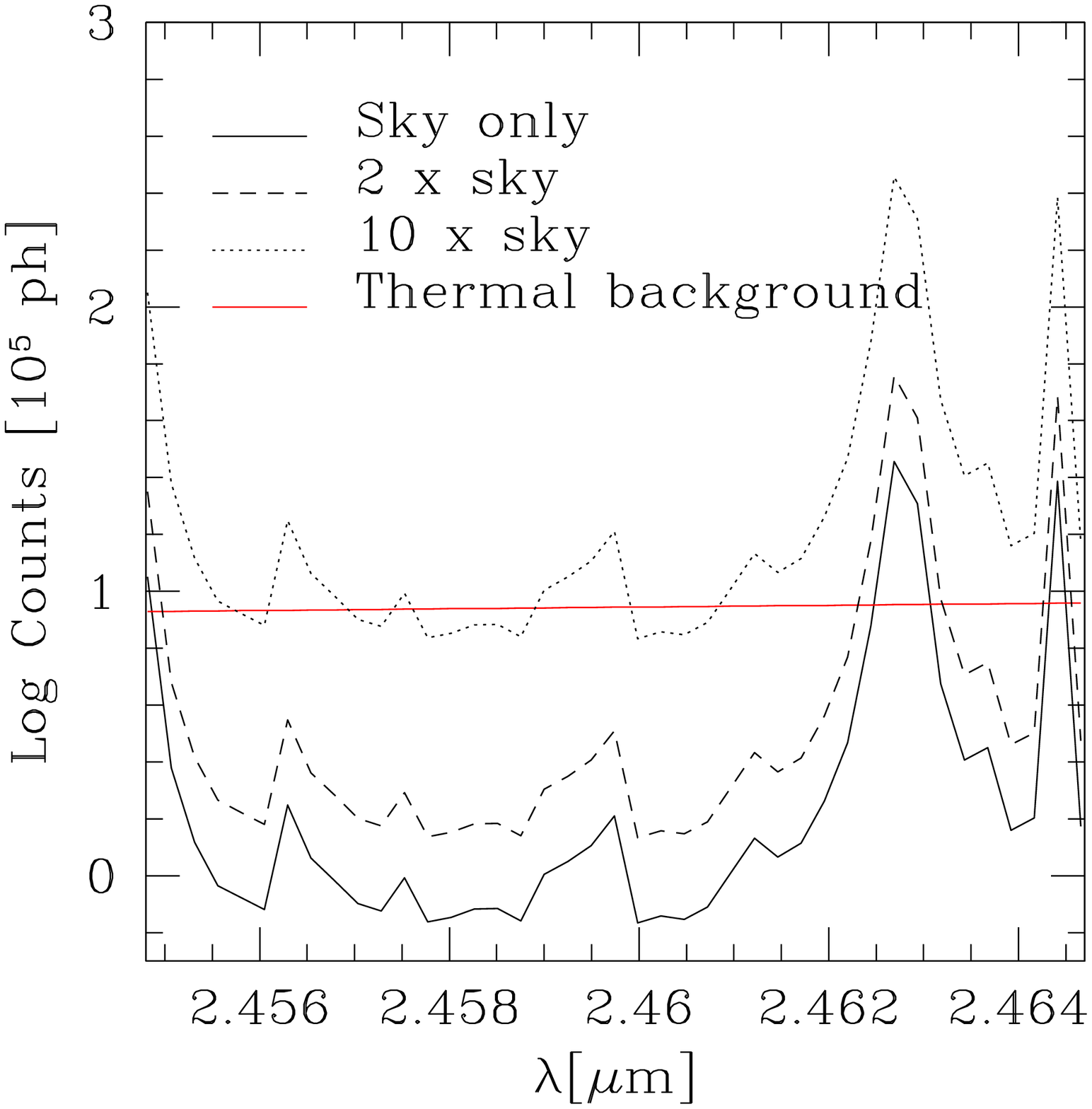}
\includegraphics[width=7cm]{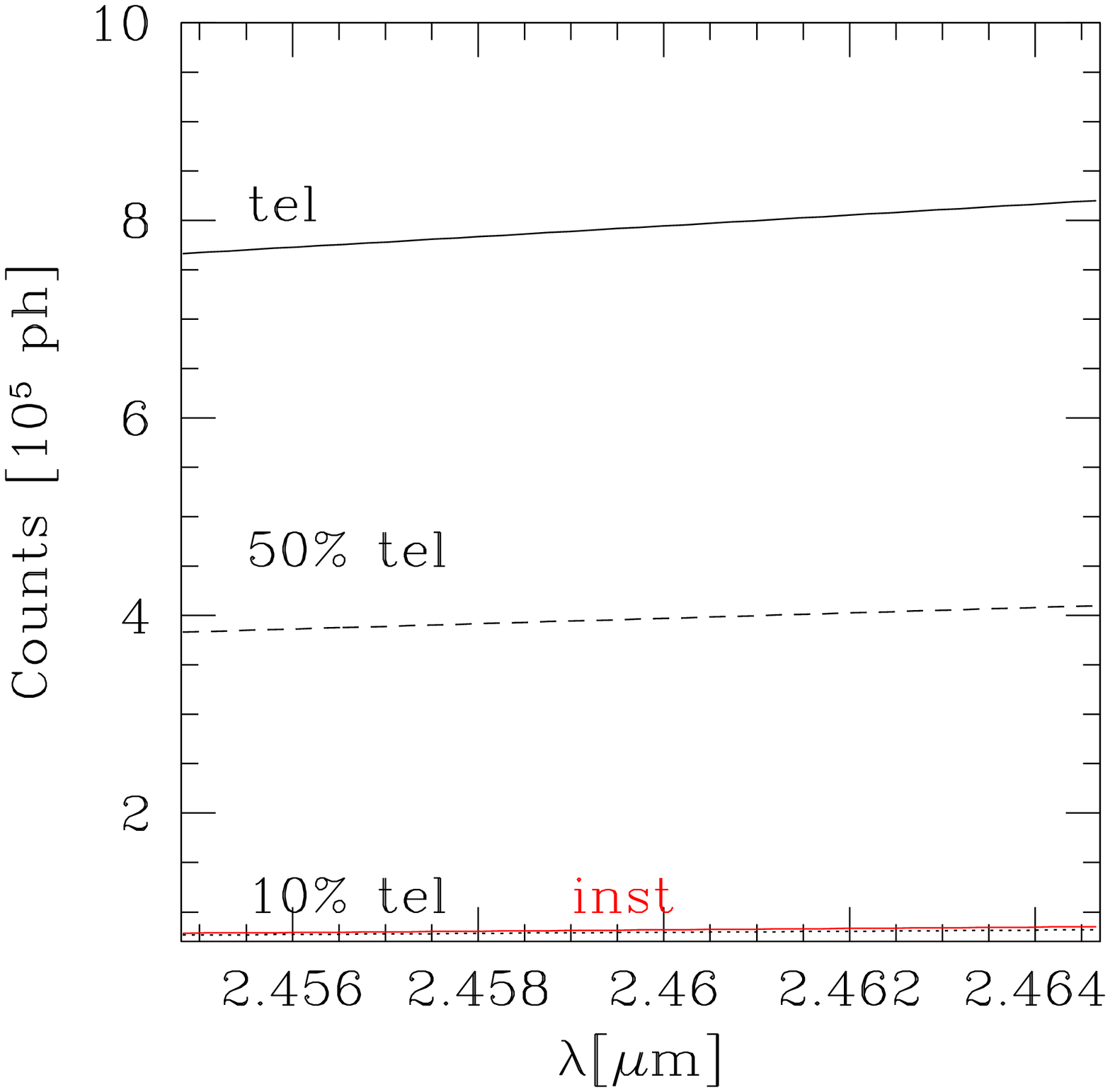}\\
\caption{Contributions to the total background in the simulations, at
z=4 (upper panel) and z=5.6 (bottom panel). \emph{Left column:}
comparisons between the sky contribution (continuum and OH sky lines,
in black) and the total thermal background (telescope and instrument,
in red). \emph{Right column:} comparison between the telescope (black)
and instrument (red) contributions to the total thermal background.}
\label{back}
\end{figure*}

{\bf Instrument model:} We assumed a reference pixel size of 50 mas,
and a reference spectral power of resolution of R=5000, as a
compromise between our desire to minimise the impact of the OH sky
lines and not wanting to over-resolve the line by a large factor. The
instrument thermal background was modelled using two gray bodies,
following a preliminary study of the thermal background of EAGLE
\citep{laporte08}. The first one models the effect of the Target
Acquisition System, and assumes a temperature of 240K and an
emissivity of 15\%. The second one models the effect of the
spectrograph, and assumes a temperature of 150K and an emissivity of
69\%. The instrument background represents less than 10\% of the
telescope background in both the H and K bands (see Fig. \ref{back}),
in agreement with the official requirements for EAGLE, which specify
that ``the instrument thermal background at the detector shall be less
than 50\% of that from the telescope'' (goal=10\%). As we will see
below, the total background is dominated by the thermal contribution
of the telescope in the K band, and by the sky background in the H
band. Therefore, we did not explore variations in the thermal
contributions from the instrument, which is never the dominant source
of background.

{\bf Detector model and exposure time:} We rely on the description of
a cooled Rockwell HAWAII-2RG IR array working at $\sim$80K, as
described by \cite{finger06}, with a read-out-noise ($RON$) of 2.3
e/pix and a dark current of 0.01 e/s/pix. Its thermal background,
bias, and saturation threshold were neglected. Because observations of
distant galaxies in the NIR are generally not limited by the detector
noise, we did not explore parameters that have a non-negligible
influence only in such a regime, i.e., variations of the individual
frame exposure time $dit$, and detector characteristics ($dark$ and
$RON$). In practice, we chose $dit$=3600s to speed up the simulations
while keeping a reasonable number of individual exposures, although
realistic exposures will be much shorter because of sky variations and
saturation of the detector. We assumed a reference case exposure time
of T$_{intg}=dit\times ndit$=24 hours, i.e., $ndit$=24.

{\bf Global throughput:} We assumed a global throughput
t$_{transm}$=20\%, detector QE included. The official DRM baseline
assumes a transmission of 90-95\% for the 5 mirror design telescope
and a QE of 90\%. The official requirements for EAGLE, specifies a
throughput of at least 35\%, including the detector QE. Therefore,
according to the baseline, the global throughput should be 31\%, which
is larger that our assumption. This will be further discussed in Sect.
5.1.1. The integrated number of photons reaching the detector is a
degenerated function of some parameters which have no impact on the
spatial or spectral resolution, i.e., $T_{intg}$, $t_{transm}$,
$EW_0$, and $D$. Therefore, we chose not to explore variations in
global transmission, which can be directly derived by analogy with
variations of associated degenerated parameters.

{\bf PSF model:} The coupling between the AO system and the 3D
spectrograph is captured through the AO system PSF. Therefore, it is a
crucial element that needs to be carefully simulated, and cannot be
approximated by, e.g., a simple Gaussian. Given the multiplex
requirements for the present science case, only GLAO \citep{rigaut02}
and MOAO \citep{assemat07} have been considered. All PSFs include the
effect of telescope aperture, but neglect the effects of the central
obscuration and spiders. They were generated at the central wavelength
of the corresponding filter (e.g., 1.65 microns for the H band), and
the difference between this wavelength and the actual wavelength where
the emission line is observed (e.g., 1.86 $\mu$m for [OII] redshifted
at z=4) was neglected.

GLAO PSFs were taken from the official DRM \emph{ftp}
depository\footnote{http://www.eso.org/sci/facilities/eelt/science/drm/tech\_data/ao/}.
These PSFs were generated by M. Le Louarn at ESO using an end-to-end
code and the official DRM baseline assumptions (v1). Of note, these
PSFs correspond to a total integration time of 4 seconds and therefore
include non-negligible speckle noise.

MOAO PSFs were simulated using an analytical code (calibrated against
an end-to-end code) by B. Neichel (GEPI-Obs. de Paris/ONERA) and T.
Fusco (ONERA), which has the advantage of producing PSFs free of
speckle noise, i.e., more representative of long-exposure PSFs
\citep{neichel08}. Briefly, we used assumptions as close as possible
to the assumptions used for GLAO PSF modelling: the pitch
(inter-actuator distance) was assumed to be 0.5m ($\sim$84$\times$84
actuators in the pupil plane), with a reference seeing of 0.8 arcsec
and an outer-scale of the turbulence $L_0$=25m; the same 10 turbulent
layer $C_n^2$ profile that the one used for GLAO PSFs was considered.
The wavefront was measured using three guide stars (assumed to be
natural guide stars, i.e., specific issues related to laser guide
stars like, e.g., the cone effect, were neglected), located on a
equilateral triangle asterism.

In order to sample different performances for the MOAO and GLAO
systems, different asterism sizes were considered. The PSFs were
systematically estimated on-axis. A detailed analysis of the coupling
between the MOAO and 3D spectroscopy can be found in \cite{puech08b}.
Of interest here is that in most cases, the coupling between the MOAO
system and the IFU pixel scale is such that the spatial resolution is
set by the latter because the PSF FWHM is smaller than twice the IFU
pixel scale. Improving the MOAO correction further does not provide
any gain in spatial resolution but still improves the Ensquared Energy
(EE) in a spatial element of resolution (i.e., the fraction of light
entering a spatial element of resolution), hence the SNR. This
justifies the usual choice of characterising the EE in an aperture
equal to twice the pixel size (see \citealt{puech08b} for details).
The characteristics of the different PSFs used in this study are listed in
Tab. \ref{tabpsf}. A detailed study of MOAO PSF shapes can be found in
\cite{neichel08}.

\begin{table}
\centering
\begin{tabular}{cc | cc}\hline
GLAO FoV(') & EE in 100 mas & MOAO FoV(') & EE in 100 mas\\\hline
1   & 15.0 & 0 & 63.6\\
2   & 12.5 & 0.3 & 56.1\\
5   & 8.2 & 0.5 & 45.6\\
10  & 6.0 & 1.0 & 33.7\\
15  & 5.3 & 2.0 & 27.2\\
 & & 3.0 & 24.1\\
 & & 4.0 & 23.1\\
 & & 5.0 & 22.7\\\hline
\end{tabular}
\caption{Sets of GLAO and MOAO PSFs generated for the simulations. The
asterism size (dubbed ``FoV'', i.e., the position of guide stars) is
varied in order to produce a range of EE sampling typical corrections.
All values were measured in the H-band with a seeing of 0.8 arcsec.}
\label{tabpsf}
\end{table}

We chose reference PSFs for both AO systems as close as possible of
the middle of their range of EE performances: the H-band MOAO
reference PSF has EE=45.6\% (in 100 mas), while the H-band GLAO
reference PSF has EE=8.2\% (in 100 mas). Unless stated otherwise,
these two PSFs are those used in the simulations.

\subsection{Simulation runs}
Unless stated otherwise, parameters are by default set to their
reference values as described above.

{\bf Influence of AO correction:}
We explore the influence of the AO correction by increasing the
EE in a given aperture as described in Sect. 3.2 (see
Tab. \ref{tabpsf}), for two types of AO systems, i.e., GLAO and MOAO.
Simulations were done for all the morpho-kinematic templates in
the MOAO case, and for the two UGC5253 and Major merger templates in
the GLAO case.

{\bf Influence of observational and physical parameters:}
Simulations were performed for five distinct stellar masses, as
described in Tab. \ref{tabmass}, and in two different runs, as
described in Tab. \ref{tabparam}.

The simulations listed in Tabs. \ref{tabmass}, \ref{tabpsf}, and
\ref{tabparam} correspond to a total of about 760 individual runs.

\begin{table*}
\centering
\begin{tabular}{|cccccc|cccccc|}\hline
RUN \#1 & & & & & & RUN \#2 & & & & &\\\hline
$z$ & $\Delta _{pix}$ & Morpho-kin. & $EW_0$ & $R$ & $D$ & $z$ & $T_{intg}$ & $\Delta _{pix}$ & AO & Morpho-kin. & Seeing\\
  &      (mas)        &   Template   &  (\AA)   &   & (m) &   &     (hr)     &       (mas)       &  correction   &  Template          & (arcsec)\\\hline
4 & 25 & UGC5253-Major merger & 30 & 5000 & 42 & 2 &  8 & 50 & MOAO & ALL & 0.8\\
  & 50 & UGC5253-Major merger & 30 & 5000 & 42 &   & 24 & 50 & GLAO & ALL & 0.8\\
  & 50 & UGC5253-Major merger & 15 & 5000 & 42 &   &    &    & MOAO & ALL & 0.8-0.95\\
  & 50 & UGC5253-Major merger & 30 & 10000 & 42 & 4 &  8 & 50 & MOAO & ALL & 0.8\\
  & 50 & UGC5253-Major merger & 30 & 2500 & 42 &   & 24 & 50 & GLAO & ALL & 0.8\\
  & 75 & UGC5253-Major merger & 30 & 5000 & 42 &   &    &    & MOAO & ALL & 0.8-0.95\\
  & 50 & UGC5253-Major merger & 30 & 5000 & 30 & 5.6 &  8 & 50 & MOAO & ALL & 0.8\\
 & & & & & &    & 24 & 50 & GLAO & ALL & 0.8\\
 & & & & & &   &    &    & MOAO & ALL & 0.8-0.95\\\hline 
\end{tabular}
\caption{Observational and physical parameters investigated in the
  first and second run of simulations. All other parameters are set to
  their reference values as described in Sect. 3.}
\label{tabparam}
\end{table*}

\section{Results of Simulations}

\subsection{Influence of AO correction}
\begin{figure}
\centering
\includegraphics[width=8cm]{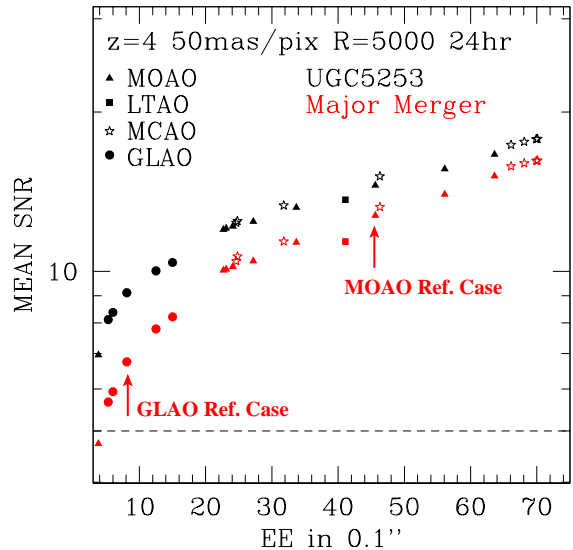}
\caption{Spatial mean SNR obtained for the reference case (H band) as
  a function of Ensquared Energy in a 100mas aperture for different
  type of AO systems. The black points correspond to a rotating disk
  template, while red points correspond to a major merger template.
  The left most symbols represent the seeing-limited case. The GLAO
  and MOAO Reference Cases (see Sect. 3.2) are indicated by red
  arrows.}
\label{aoee}
\end{figure}

The spatial-mean SNR obtained for the reference case (H band) as a
function of Ensquared Energy in a 100 mas aperture (i.e., twice the
reference pixel size, see above) is plotted in Fig. \ref{aoee}. For
comparison, we also plotted results using the major merger template.
This figure shows that, all simulations approximately fall along the
same curve, independently of the AO type, although different codes
(end-to-end vs. analytical) were used (see Sect. 3.2). This means
that, considering a given morpho-kinematic template, all the different
PSFs can be safely compared. For comparison to the GLAO and MAO PSFs
described in Sect. 3.2, we have added MCAO PSFs (with 3DMs and
asterisms of 0.5 and 5 arcmin and corrections derived at 0, 0.5,2, and
2.5 arcmin away from the centre of the FoV) and LTAO PSFs (FoV of 45
arcsec, on axis). Details about how these PSFs were generated can be
found in \cite{neichel08b}. The PSFs picked up as representative
performances of the GLAO and MOAO modes (see Sect. 3.2) have been
indicated by red arrows. In the H band, they can be considered as well
representative of the AO performance within a range of $\sim \pm$2 in
SNR for MOAO, and $\sim \pm$3 for GLAO. A systematic study of the
impact of EE on the kinematics of distant galaxies using MOAO can be
found in \cite{puech08b}. Of interest here is that the reference MOAO
PSF, with an EE of 45.6\% in an aperture of 100 mas is above the
minimal requirements derived in this study.

\subsection{DRM Steps}

In the rest of this section, we investigate the impact of the
instrument, site, and telescope, on the scientific capabilities of the
E-ELT equipped with a NIR-IFS in each of the DRM goals/steps as
defined in Sect. 2.1.

\subsubsection{DRM STEP 1: 3D detection}

\begin{figure*}
\centering
\includegraphics[width=5.8cm]{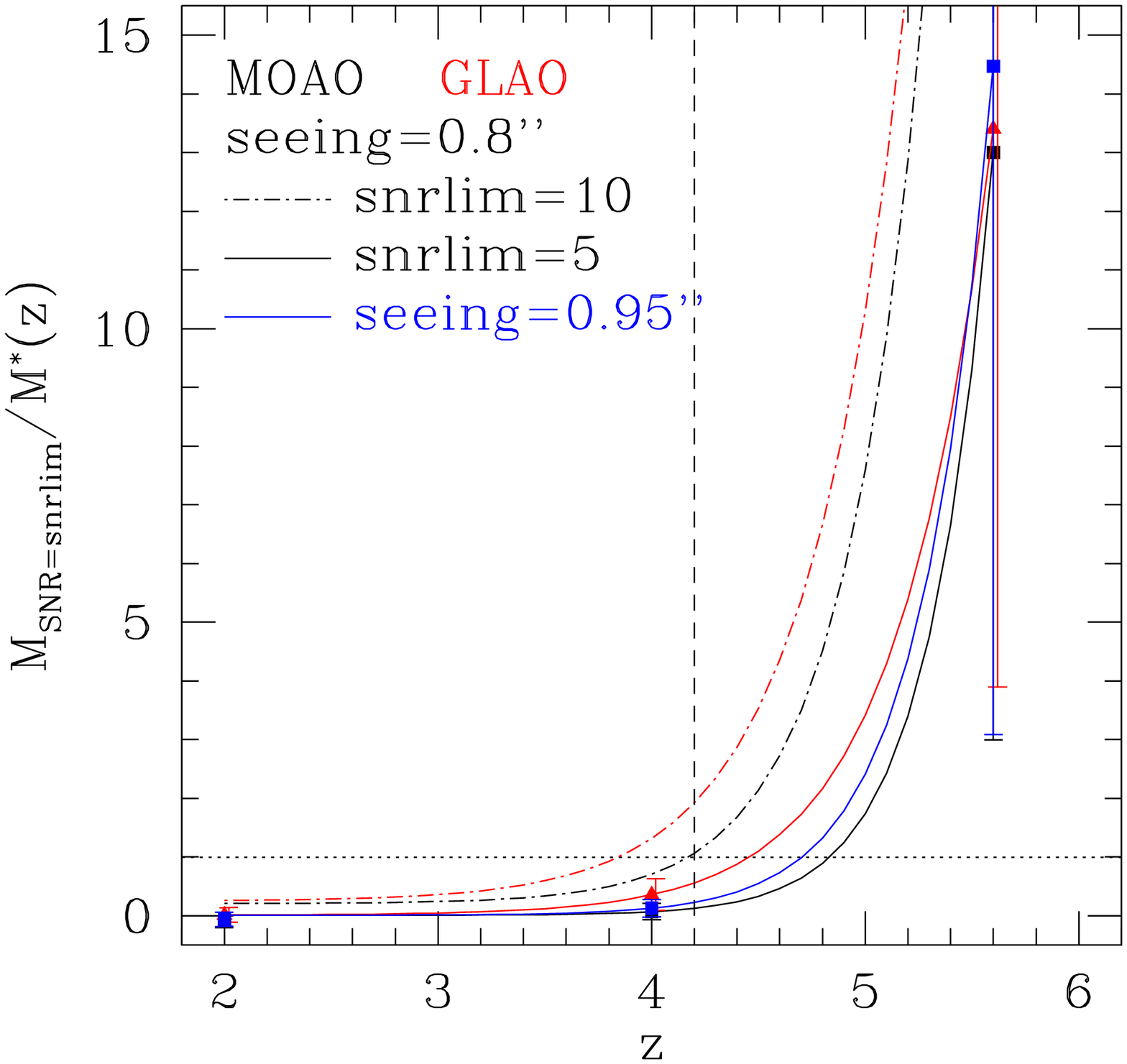}
\includegraphics[width=5.8cm]{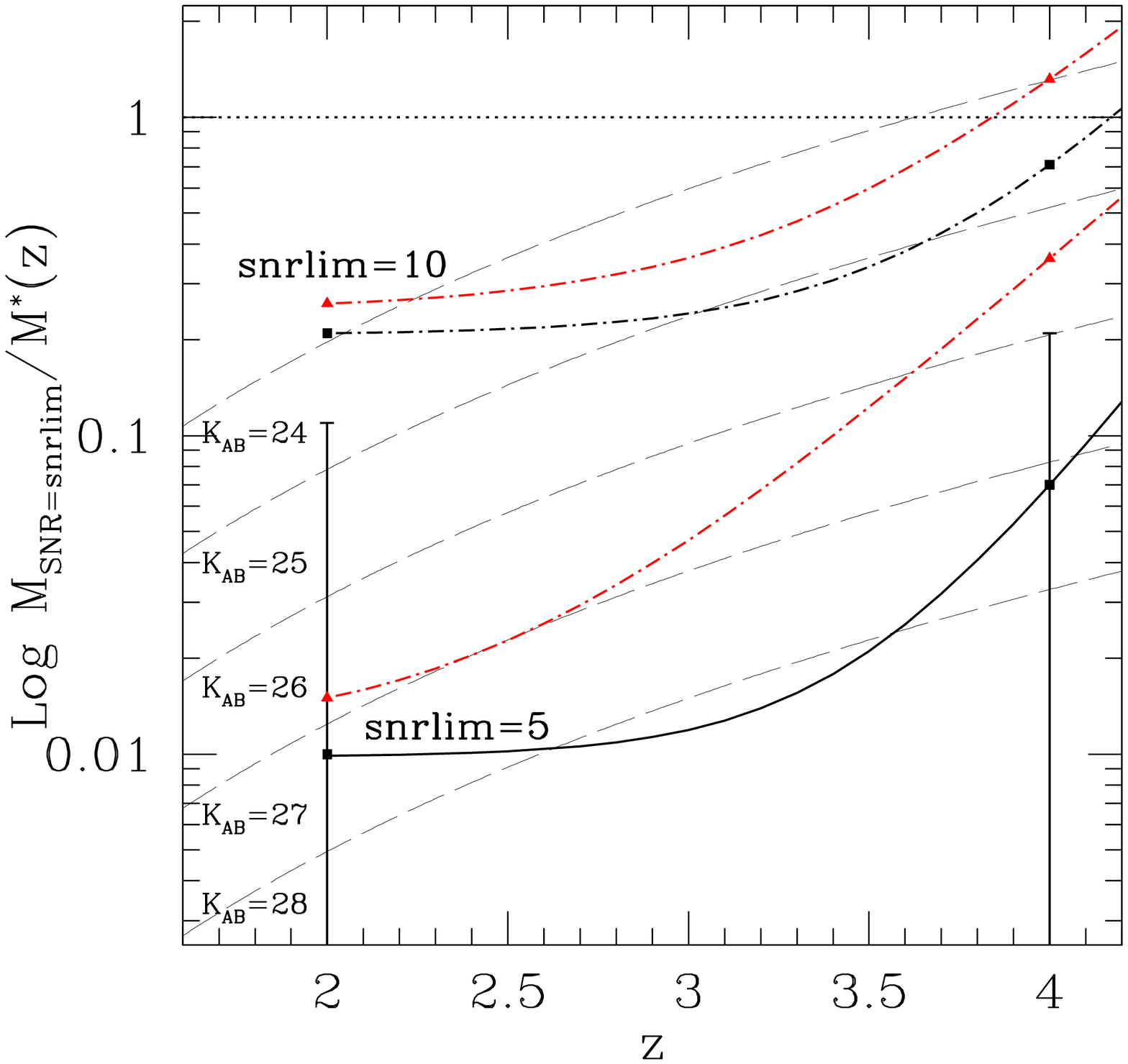}
\includegraphics[width=5.8cm]{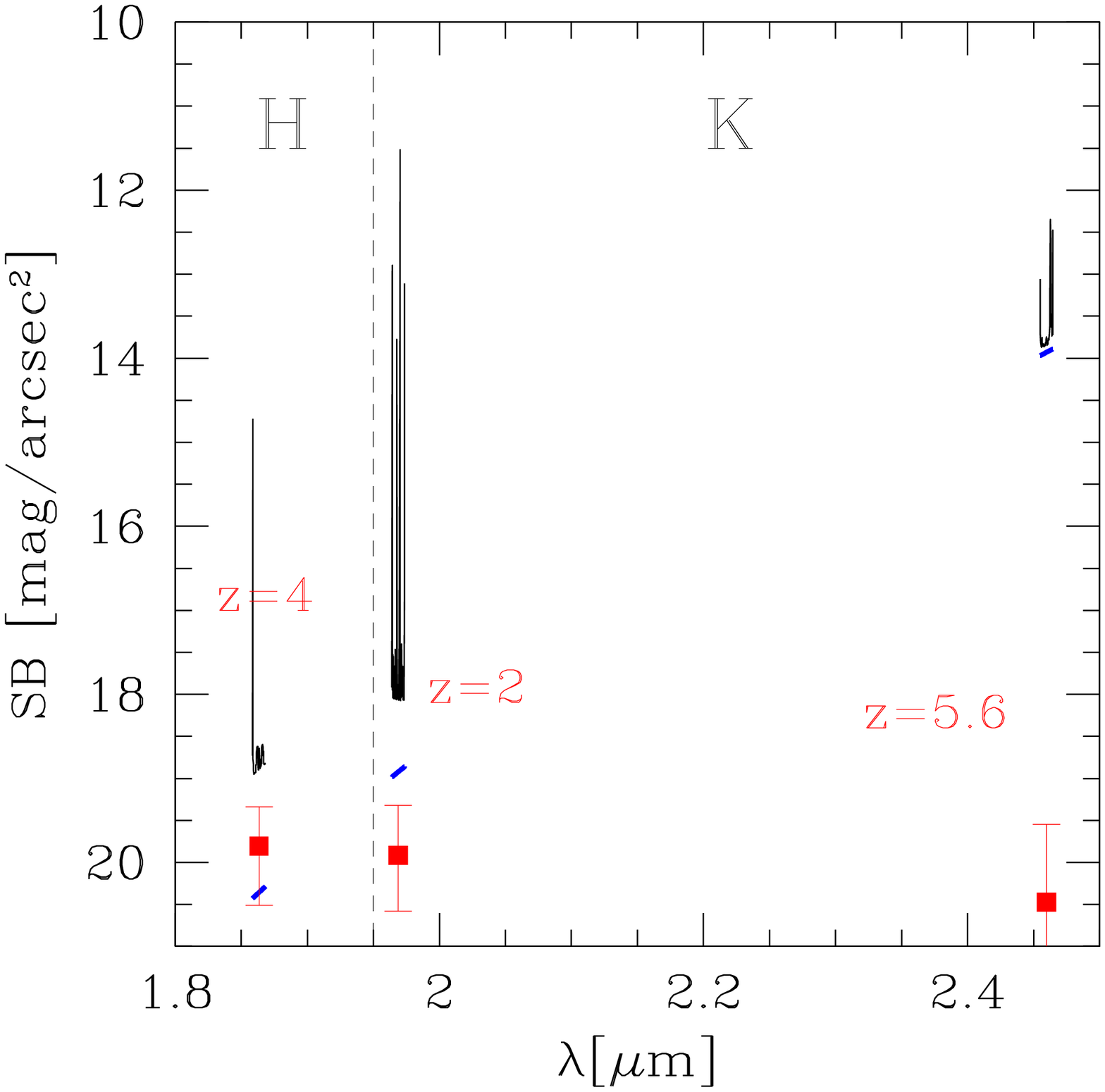}\\
\caption{\emph{Left:} Stellar mass limit that can be reach at a given
  $SNR_{min}$ as a function of redshift, AO correction, and seeing.
  Error-bars represent the range of mass limit obtained when
  considering all the morpho-kinematic templates. The vertical dashed
  line shows the limit above which the [OII] emission line gets
  redshifted into the K band. \emph{Middle:} Zoom of the redshift
  range [2,4]. Long-dashed lines show iso-magnitude curves. For
  clarity, only a subset of all the curves shown in the main panel are
  reproduced here, and the range of mass limit is shown only for the
  MOAO cases with 0.8 arcsec seeing. \emph{Right:} Comparison between
  the median emission line surface brightness of the simulated
  galaxies (in red; the error-bars represent the values found in the
  0.1-10M$*$ range) with that from the thermal background (telescope
  and instrument, blue lines) and total background (thermal plus
  continuum and OH sky lines, black lines). For saving computation
  time, only narrow spectral ranges around the targeted emission lines
  were simulated for the three redshift cases.}
\label{masslim}
\end{figure*}

We adopt a lower limit in $SNR$ of $SNR_{min}=5$ (spatial-mean SNR in
the emission line, see Sect. 2.3) for the 3D detection of distant
galaxies. The SNR obtained as a function of the observational parameters
and mass was studied in details in \cite{puech08c}. In this study, we
found that the minimal SNR scales as follows:

$$
SNR_{min}=5\times\bigg(\frac{T_{intg}}{24h}\bigg)^{0.5}\bigg(\frac{D}{42m}\bigg)\bigg(\frac{EW_0}{30\AA}\bigg)\bigg(\frac{R}{5000}\bigg)^{-0.5}\bigg(\frac{\Delta _{pix}}{50mas}\bigg)
$$

This scaling corresponds to what is expected in a background-limited
regime, and applies to all morpho-kinematic templates. As discussed in
Sect. 3.2, $T_{intg}$ and the total throughput (not explicitly
explored by simulations) follow the same scaling.

Figure \ref{masslim} gives the stellar mass that can be reached as a
function of redshift for MOAO corrections, a seeing of 0.8 arcsec, and
$SNR_{min}=5$ (black line). In this plot, error-bars represent the
range of stellar mass derived considering all the morpho-kinematic
templates. We adopted the middle of this range as a simple estimator
of the stellar mass one can reach at a given redshift under the
different specific observational conditions considered in this plot
(see Fig. \ref{masslim}). One can see that the limit in stellar mass
grows exponentially with redshift:
$$
\frac{M_{lim}}{M_*}\simeq 3.3\times 10^{-6}\exp{\bigg(\frac{z}{0.37}\bigg)},
$$which is due to the exponentially increasing contribution of the
thermal background from the telescope in the NIR (see right panel in
Fig. \ref{masslim}). Noteworthy, the mass limit evolves quite smoothly
with redshift up to z$\sim$4-4.5: this means that this limit does not
depend very strongly on input parameters like, e.g., seeing (see the
blue curve for a 0.95 arcsec), or AO correction (see the red curve for
GLAO corrections). It is not surprising that MOAO and GLAO give
similar results, as the SNR considered here is a spatial-mean over the
galaxy diameter at constant spatial and spectral sampling, which does
not take into account differences in terms of spatial resolution (see
next Section). One can adopt $SNR_{min}=10$ (instead of 5) without
impacting too strongly the resulting mass limit.

It is useful to summarise Fig. \ref{masslim} by deriving the redshift
up to which the Galaxy Stellar Mass Function (GSMF) can be probed down
to M$_*$, as galaxies having such a stellar mass at a given $z$ are
those contributing the most to the stellar mass density at this
redshift. From Fig. \ref{masslim}, one can see that \emph{the GSMF can
be probed down to M$_*$ up to z$\sim$4, independently of the SNR limit, AO
correction, and seeing conditions}.

\subsubsection{DRM STEP 2: Large scale motions}
As detailed in \cite{puech08b}, large scale motions carry important
information about the dynamical nature of galaxies. In this section,
we focus on the two extreme cases of morpho-kinematic variations found
amongst the galactic zoo, which are regular rotating disks as opposed
to major mergers. Indeed, because the latter are the most violent
galactic events, they are expected to result in the largest
morpho-kinematic amplitudes. In this section, we consider only the
UGC5253 and the major merger templates to illustrate how large-scale
motions can be used to distinguish between a rotating disk and a major
merger using spatially-resolved kinematics on the E-ELT.

It is difficult to build a simple criterion that would allow us to
assess the quality of the recovered large-scale motions, and quantify
how well one can distinguish between a rotating disk and a merger in
any situation. At first order, such a criterion is a function of the
number of spatial element of resolutions available, and of the surface
brightness detection limit. Present methods to distinguish between
rotating disks and mergers are indeed limited to specific range of SNR
and/or spatial resolution (see, e.g., \citealt{shapiro08, flores06}).
Finding a method for analysing the velocity fields of distant galaxies
with a large range of spatial resolution and/or SNR in a uniform way
will be challenging and is far beyond the scope of this study.

Here, we adopted a simple criterion based on a threshold on the
spatial-mean SNR over the intrinsic galaxy diameter, as defined in
Sect. 2.3. Using MOAO, it has been suggested that a spatial-mean SNR
of 5 is a minimum to recover large-scale motions and distinguish
between a major merger and a rotating disk \citep{puech08b}. It is
clear that such a criterion is generally too simplistic to allow us to
distinguish between a merger and rotating disk in any possible case
encountered in nature (see discussion in \citealt{puech08b}). However,
it captures the two required basic dependencies on surface brightness
detection (i.e., only pixels having a kinematic $SNR_{kin}$ larger
than 3 are considered for the kinematic analysis, see Sect. 2.3) and
on the number of available element of spatial resolutions (i.e., the
spatial-mean SNR is derived over the intrinsic galaxy diameter
independent of what parts are detected, see Sect. 2.3). Observations
of z$\sim$0.6 galaxies using FLAMES/GIRAFFE at the VLT have
demonstrated that only 3 spatial element of resolution can already
allow us to distinguish between a rotating disk and more complex
systems \citep{flores06,yang08}. We checked that in 98\% of the
simulations, a spatial-mean SNR of at least 5 corresponds to at least
3 spatial elements of resolution. Therefore, this simple criterion
provides us with a useful guidance on data quality for kinematic
classification, well within the scope of the present study.

\begin{figure*}
\centering
\includegraphics[angle=90,width=12.5cm]{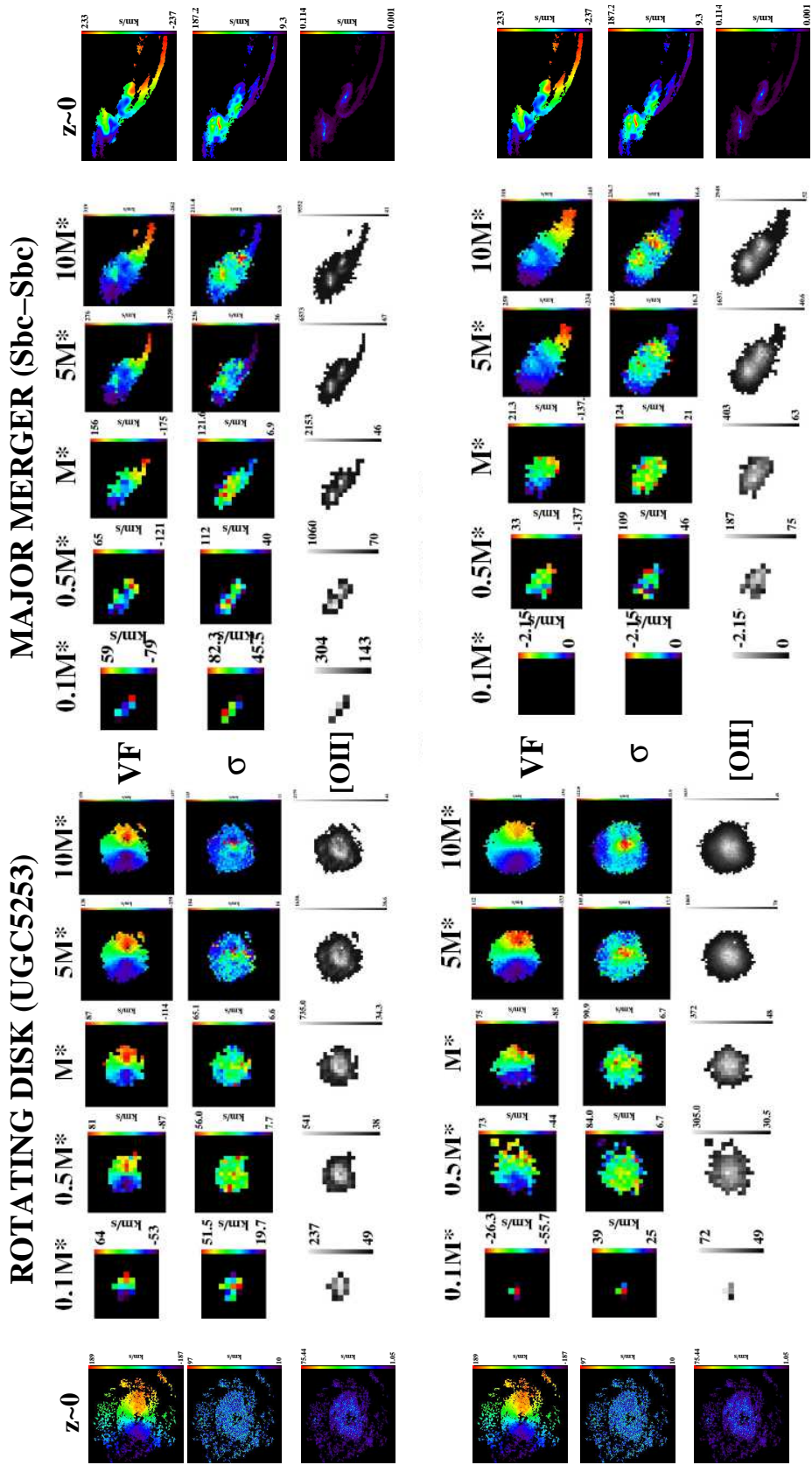}
\caption{Results of simulations at z=4 with MOAO for the reference
  case (upper panels; the size of the stamps is, from left to right,
  0.55, 0.95, 1.2, 2.1, and 2.7 arcsec for the rotating disk template,
  and 0.4, 0.7, 0.9, 1.1, and 2.0 arcsec for the major merger
  template) and GLAO (bottom panels). Each panel shows (\emph{from top
    to bottom} the velocity field (first sub-panels), the velocity
  dispersion map (second sub-panels), and the emission line map (third
  sub-panels) for different stellar masses (from 0.1M$_*$, first
  column, to 10M$_*$, last column). The two panels on the left show
  simulations using a rotating disk template (UGC5253), while the two
  panels of the right show simulations for a major merger. On the
  leftest and rightest sides, we have reproduced the input templates
  shown in Fig. \ref{templates} to ease the comparison with
  simulations, although the spatial scales are different.}
\label{sim}
\end{figure*}

In Table \ref{tabsnr}, we give the SNR obtained for a range of
stellar-mass using the reference case and MOAO or GLAO corrections.
According to this table, it should be feasible to distinguish between
both kinds of templates down to 0.5M$_*$ at z=4 using MOAO. A direct
visual inspection of the simulations confirms this threshold (see Fig.
\ref{sim}).

\begin{table}
\centering
\begin{tabular}{ccc}\hline
M$_{stellar}$ & SNR w/ MOAO & SNR w/ GLAO\\
(in $M_*(z)$) &      z=4 &        z=4\\\hline
0.1  & 5.45-3.85   & 2.58-2.53   \\
0.5  & 11.09-8.80  & 6.41-4.40   \\
1.0  & 15.54-12.76 & 9.12-6.75   \\
5.0  & 24.84-26.07 & 17.89-16.46 \\
10.0 & 32.24-33.27 & 24.21-23.46 \\\hline
M$_{stellar}$  & SNR w/ MOAO & SNR w/ GLAO\\
(in $M_*(z)$) &    z=5.6 &     z=5.6\\\hline
0.1 & 0.78-0.44 & 0.37-0.18\\
0.5 & 1.49-1.23 & 0.90-0.60\\
1.0 & 2.07-1.60 & 1.28-0.86\\
5.0 & 3.92-3.75 & 2.76-2.20\\
10.0 & 5.14-5.13 & 3.66-3.15\\\hline
\end{tabular}
\caption{Spatial-mean SNR obtained for simulations of the reference
  case as a function of mass and redshift, for MOAO (upper part) and
  GLAO corrections (bottom part). The first value corresponds to
  simulations with a rotating disk template (UGC5253), while the
  second value corresponds to simulations with a major merger
  template.}
\label{tabsnr}
\end{table}

Using GLAO, Table \ref{tabsnr} suggests that the same distinction
could be made down to M$_*$ at z=4 (instead of 0.5M$_*$ using MOAO).
Since GLAO corrections smooth out many more kinematic details than
MOAO, it is difficult to distinguish a major merger from a rotating
disk down to the same limit in stellar mass/SNR: it is indeed
difficult to visually distinguish non-circular motions in the 0.5M$_*$
major merger simulations at z=4 with GLAO, as shown in Fig. \ref{sim}.

\begin{figure}
\centering
\includegraphics[width=8.5cm]{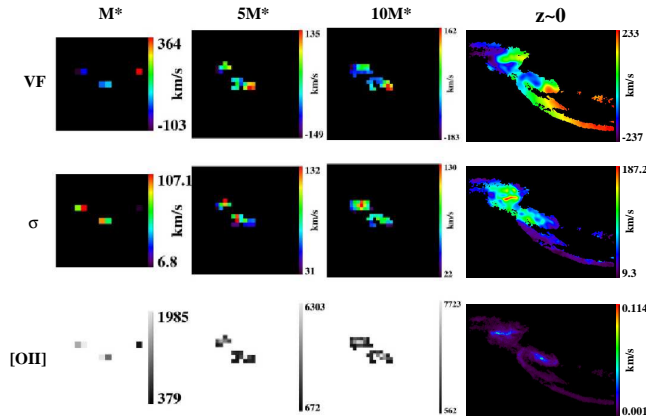}
\caption{Results of major merger simulations at z=5.6 using MOAO.
  \emph{From top to bottom:} velocity fields (first panel), velocity
  dispersion maps (second panel), and emission line maps (third panel)
  for different stellar masses (from 1M$_*$, first column, to 10M$_*$,
  last column). Lower mass systems are not detected. The size of the
  stamps is, from left to right: 0.71, 1.35, and 1.7 arcsec. On the
  rightest side, we have reproduced the input templates shown in Fig.
  \ref{templates} to ease the comparison with simulations, although
  the spatial scales are different.}
\label{sim5.6}
\end{figure}

At z=5.6, the much higher background makes any distinction very
difficult, and at best limits this exercise to the highest mass
galaxies, as suggested by Tab. \ref{tabsnr} and Fig. \ref{sim5.6}.

In summary, \emph{provided a SNR of 5, these simulations suggest that
  it is possible, using MOAO, to recover large scale motions of
  galaxies and distinguish between different dynamical states at least
  up to z=4 and down to M$_{stellar}$=0.5M$_*$. GLAO appears to
  provide very similar performances, although at lower spatial
  resolution, which would limit the recovery of the dynamical state
  down to $\sim$M$_*$ at z=4. At higher redshift, the loss of SNR
  induced by the increasing contribution of the thermal background of
  the telescope, will limit the recovery of large-scale motions to very
  massive systems.}

\subsubsection{DRM STEP 3: Rotation Curves}
\begin{figure*}
\centering
\includegraphics[width=8.5cm]{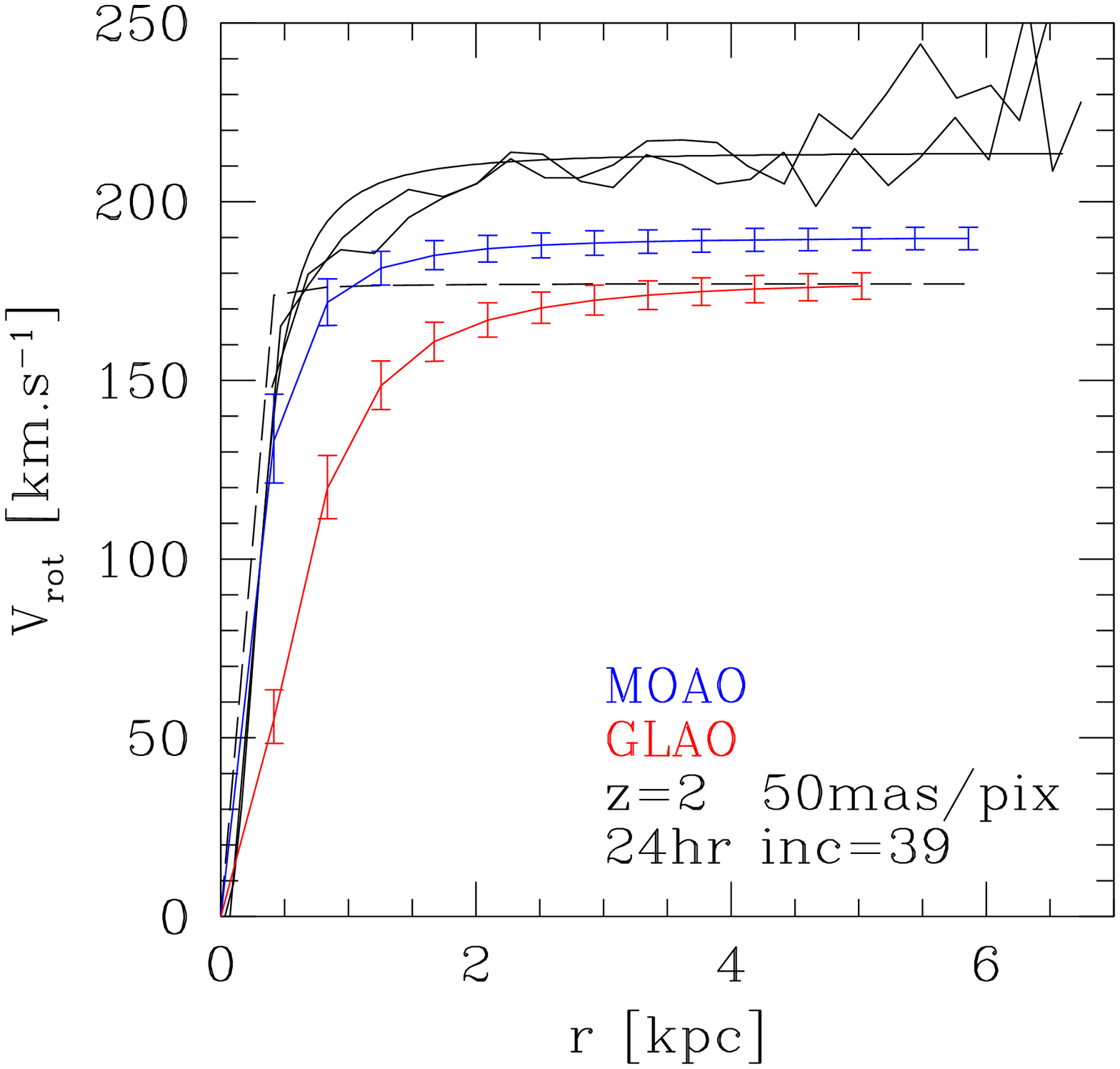}
\includegraphics[width=8.5cm]{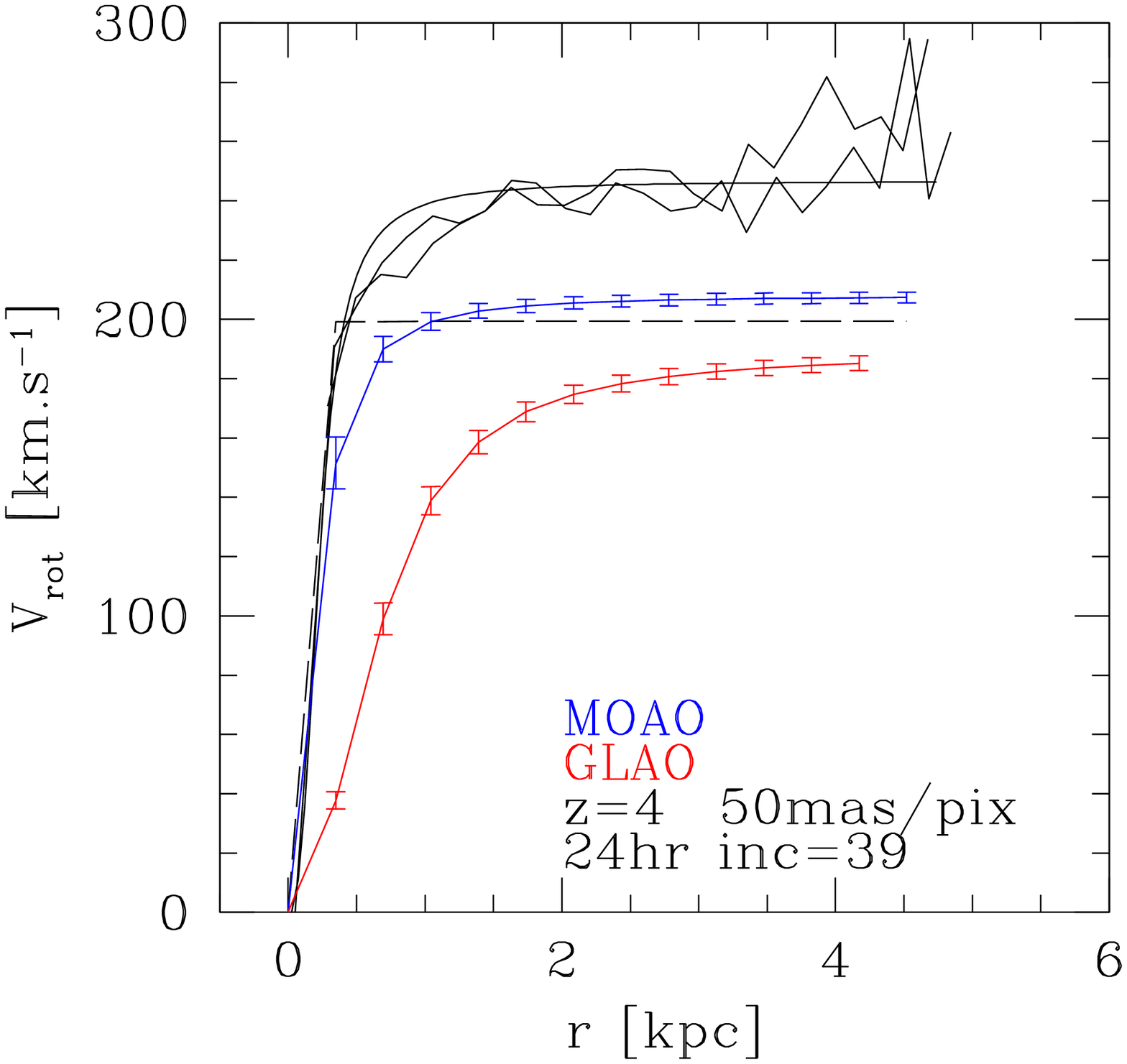}
\caption{Rotation curve of a $z$=2, $M_*$ galaxy (left), and a $z$=4,
  5$M_*$ galaxy (right). The black lines represent the original z=0
  rotation curve rescaled to the size of simulated z=2 and z=4
  galaxies. The rotation curve on each side of the galaxy is shown,
  along with an arctan fit. The rotation curve obtained using MOAO is
  shown in blue, while the one obtained using GLAO is shown in red.
  Error-bars represent only the uncertainty associated to the two
  parameters of the arctan function, and do not take into account
  uncertainties associated to other kinematic parameters like the
  dynamical centre or the Principle Axis. The black dashed line
  represent the rotation curve obtained in simulations without PSF:
  they correspond to the best rotation curve that one can obtain at
  the pixel scale of the simulations (50mas/pix).}
\label{rc}
\end{figure*}

In Figure \ref{rc}, we show rotation curves extracted from simulations
at z=2 and z=4 using the UGC5253 rotating disk template. This template
has the steepest velocity curve gradient among the morpho-kinematic
templates, and is therefore useful to assess at which extent which
parts of the rotation curve can be recovered safely. Indeed, the
central part of the rotation curve is well-known to be strongly
affected by limited spatial resolution, an effect known for long as
``beam smearing'' by HI observers. Among the dynamical parameters
fitted to derive a rotation curve from a velocity field (see, e.g.,
\citealt{epinat08} for details), inclination is by far the one that
carries the largest uncertainty. Indeed, there is a well known
degeneracy between inclination and rotation velocity during such a
fitting. \cite{epinat09b} have extensively studied these degeneracies
as a function of beam smearing using artificially redshifted galaxies
at z=1.7. These authors concluded that the best strategy is to use
high-resolution broad-band imaging to derive a morphological estimate
of the inclination, and use this to relax one parameter during the
kinematic model fitting. We therefore did not try to fit the
inclination, which was held constant during the velocity field, and
focused instead on the influence of different AO modes in recovering
spatial features of the rotation curve. In other words, we assumed
that high-resolution imaging will be available to provide the observer
with unbiased estimated of this parameter. We adopted a simple arctan
rotation curve model, which is fully described by two parameters
\citep{courteau97}.

At z=2, only MOAO can more or less recover the rising part of the
rotation curve (RC). Due to its much lower spatial resolution
(FWHM$_{GLAO}$=161mas and FWHM$_{MOAO}$=11mas), GLAO induces a strong
beam smearing-like effect, as in HI observations of local galaxies.
Even with MOAO, the spatial sampling of the IFU (50mas/pix) limits the
spatial resolution of the observations (to $\sim$0.1 arcsec) and does
not allow us to recover the true shape of the RC. Compared to the best
RC that one can recover at this spatial scale (see the black dashed
line), MOAO and even GLAO do not induce any bias in terms of rotation
velocity. However, numerical simulations will be needed to recover the
true rotation velocity, as it is already the case in lower redshift
kinematic studies of distant galaxies (see, e.g., \citealt{flores06};
\citealt{forster06}; \citealt{puech08}).

At z=4, the lack of spatial resolution will limit the study of
rotation curves to super-M$_*$ rotating disks, as shown in Fig.
\ref{rc}. At these redshifts, only MOAO will provide enough spatial
resolution to limit biases in recovering the rotation velocity or the
rising part of the RC.

In summary, \emph{the recovery of the whole shape of the RC will
  propably be limited to z$\sim$2, using MOAO. The recovery of the
  true rotation velocity will require numerical simulations, which is
  already the case in lower-z studies. At higher redshifts, such
  measurements will probably be limited to super-M$_*$ galaxies. Using
  GLAO strongly degrades the shape of the recovered RC and probably
  prevents any accurate recovery of its true shape.}

\subsubsection{DRM STEP 4: Detailed kinematics}
\begin{figure*}
\centering
\includegraphics[width=18cm]{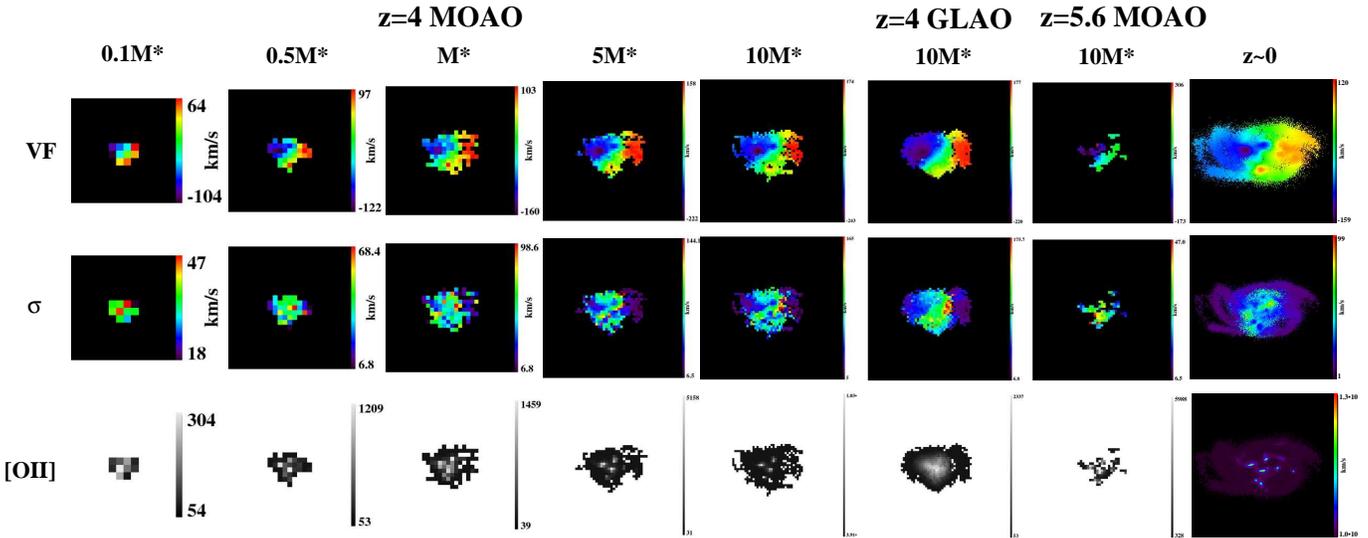}
\caption{\emph{From left to right:} Simulations of clumpy disks at z=4
  using MOAO (from 0.1M$_*$, first column, to 10M$_*$, last column;
  the size of the stamps is, from left to right: 0.7, 1.25, 1.6, 2.8,
  3.55 arcsec), GLAO at z=4 (10M$*$; the size of the stamp is 3.55
  arcsec), and MOAO at z=5.6 (10M$*$; the size of the stamp is 3.0
  arcsec). For each column, the first line shows the velocity fields,
  the second line the velocity dispersion maps, and the third line the
  emission line maps. On the rightest side, we have reproduced the
  input templates shown in Fig. \ref{templates2} to ease the
  comparison with simulations, although the spatial scales are
  different.}
\label{clumps}
\end{figure*}

To illustrate the capability of the E-ELT in recovering the detailed
kinematics of distant galaxies, we consider in this section the result
of simulations with the Jeans-unstable clumpy disk templates shown in
Fig. \ref{clumps}. Clumps can clearly be distinguished in galaxies
more massive than M$_*$ at z=4, using MOAO (see emission line maps).
Using GLAO does not allow us to identify these clumps anymore: even in
the most massive galaxy simulated, all clumps are smoothed together in
the emission line map. One can still recover non-circular motions in
the velocity field, but without clear morphological signature, it is
difficult to identify the underlying cause of this perturbation. At
higher redshift, the limited SNR does not allow us to recover such
clumps anymore, even in the most massive case.

In summary, \emph{these simulations suggest that it is possible to
  recover clumps in rotating disks down to z=4, M$_*$ galaxies using
  MOAO. Using GLAO probably prevent any direct identification of
  clumps in very distant galaxies.}

\subsection{Compliance with figures of merit}
The goal of this simulated experiment was to study a large number
$N_{gal}$ of galaxies (in order to allow one to derive statistics
over the morpho-kinematic types within each mass and redshift bin) at
$2\leq z< 6$ with $0.1 \leq M_{stellar} \leq 5\times 10^{11}
M_\odot$ in less than 100 nights. We adopt the following assumptions
as a typical observational strategy for this program:
\begin{itemize}
\item Reference case with R=5000 and $\Delta _{pix}$=50mas (see Sect.
3);

\item MOAO corrections;

\item Mauna-Kea-like background, as described in Sect. 3;

\item A limiting SNR of 10 (10-$\sigma$ detection);

\item Overheads of 30\% (i.e., overhead factor $OH$=1.3);

\item The number of effective observed hours per night is assumed to
be 8 hr;

\item Three redshift ``bins'' ($z$=2, $z$=4, and $z$=5.6) and three
  mass ``bins'' per redshift bin ($\sim$0.5$M_*$,$\sim M_*$, and
  $\sim$5$M_*$), except for the $z=5.6$ case, which has only two mass
  bins ($\sim M_*$, and $\sim$5$M_*$), since we are interested in
  galaxies having $0.1 \leq M_{stellar} \leq 5\times 10^{11} M_\odot$
  and that M$_*$(z=5.6)=0.8$\times$10$^{10}$M$_\odot$. We therefore
  assume that the survey is divided into $N_{bins}$=8 elementary bins;

\item The number of galaxies per elementary bin is assumed to be
  $N_{gal/bin}$. This translates into total number of galaxies in the
  survey of $N_{gal/bin} \times N_{bins}$.

\item Within each bin of the survey, the $N_{gal/bin}$ galaxies can be
  observed in $N_{setups}$ setups, with $N_{setups}$=$N_{gal/bin}$/$M$
  where $M$ is the multiplex advantage of the instrument, i.e., the
  number of galaxies that can be observed at the same time. We assume
  $M \geq N_{gal/bin}$, which means that all galaxies in a given bin
  can be observed using only one instrument setup ($N_{setup}=1$).
\end{itemize}

Under these assumptions, the corresponding integration times, in
nights, needed in each elementary bin are given in Tab.
\ref{tabsurvey}. The great collecting power of the E-ELT is clearly
reflected in the time needed to survey the redshift range 2-4, which
requires only $\sim$7 nights. Regarding the z=5.6 bin, the large
thermal background of the telescope (see Figs. \ref{back} and
\ref{masslim}) translates into very large integration times.

\begin{table}
\centering
\begin{tabular}{cccc|c}\hline
T$_{intg}$ & 0.5M$_*$ & M$_*$ & 5M$_*$ & Total\\\hline
z=2   & 1.2 & 0.8 & 0.3 & 2.3\\
z=4   & 2.3 & 1.4 & 0.6 & 4.3\\
z=5.6 & --- & 66.9 & 16.39 & 83.2\\\hline
Total & 3.5 & 69.1 &  17.2 & 89.8\\\hline
\end{tabular}
\caption{Integration time needed to complete the survey, in nights,
  assuming a multiplex $M \geq N_{gal/bin}$.}
\label{tabsurvey}
\end{table}

The total time of the survey, in nights, can be expressed as follows:
$$
T_{survey}(n)\simeq 90\bigg(\frac{SNR _{lim}}{10}\bigg)^2\bigg(\frac{OH}{1.3}\bigg)\bigg(\frac{N_{gal}}{8M}\bigg)
$$

As an example, if one wants to construct a 90-nights survey with a
very large sample of $N_{gal}$=1000 galaxies, a multiplex factor of
$M$=125 will be required. However, it is likely that the future
multi-object integral field spectrograph on the E-ELT, such as EAGLE
\citep{cuby08}, will not provide us with such a large multiplex
capability. At constant $T_{survey}$, the relation between the
multiplex capability $M$ and the total number of galaxies in the
survey is $N_{gal}=8M$. Therefore, if one still requires such a large
number of galaxies with a lower multiplex factor, e.g., $M$=25, then a
total observation time of 5$\times$90=450 nights will be necessary to
complete the survey. Alternatively, one can decide to reduce the
number of galaxies: with $M$=25, 200 galaxies can be observed within
$T_{survey}$=90n. Such a program would certainly provide a very
interesting and useful first glimpse of the galaxy mass assembly
process as a function of time.

In summary, \emph{with a total integration time of 90 nights, and
  within the assumptions listed above, it will be possible to:}
\begin{itemize}
\item \emph{Observe $N_{gal}$ galaxies with SNR=10 in three redshift
  bins (2-4-5.6) and three mass bins $(0.5M_*$-$1.0M_*$-$5M_*)$,
  except for the $z=5.6$ bin, which has two mass bins ($\sim M_*$, and
  $\sim$5$M_*$); all bins have $M$ galaxies (i.e., the multiplex gain
  of the instrument), which will allow one to do statistics over
  morpho-kinematic types;}

\item \emph{Recover the large-scale motions of galaxies at least up to z=4
and down to $M_{stellar}=0.5M_*$, i.e., of most galaxies in the
survey;}

\item \emph{Recover the detailed kinematics of the most massive galaxies
(the detection of clumps in rotating disks will be possible down to z=4,
M$_*$ galaxies).}
\end{itemize}

\section{Discussion}

\subsection{Limitations and accuracy of simulations}

\subsubsection{Observational inputs}
For simplicity, we adopted several assumptions that might impact the
results of the simulations in terms of SNR (see Sect. 3). First, we
adopted a conservative global throughput, while proposed designed
(e.g., EAGLE) predict throughput higher by a factor $\sim$55\%. This
would translate into an increase of the SNR by $\sim$25\%, since $SNR
\propto \sqrt{t_{trans}}$. Second, we have neglected the central
obscuration of the E-ELT, which leads to overestimate the SNR by
$\sim$8\%, since $SNR \propto D$. Finally, we have assumed that the
[OII] doublet in a single emission line instead of a doublet. The
average [OII] line ratio is 1.4 \citep{flores06}, which leads to an
overestimation of the maximal SNR in the emission line by a 41\% at
constant flux (based on the crude assumption of two resolved
Gaussians), which in turn leads to overestimate the spatial-mean SNR
by the same factor. All these assumptions lead to overestimate the
spatial mean SNR by $\sim$24\%, or, equivalently, the simulations
correspond to an $EW_0 \sim$37\AA~instead of 30\AA, which was chosen
as an extrapolation of the \emph{mean} $EW_0$([OII]) observed at
z$\sim$1 \citep{hammer97}. This new value is actually closer to the
\emph{median} $EW_0$([OII]) observed at z$\sim$1 \citep{hammer97}:
such a value is as relevant as the first one, and therefore we
conclude that the adopted simplifications do not impact strongly the
results.

The most important limiting factors related to the observational
parameter space are certainly the emissivity of the telescope, which
drives the thermal emission from the telescope, which limits the
detectability of high-z galaxies in the K band, and the brightness of
the sky continuum between OH lines, which limits observations at lower
redshift (see Sect. 4.2). As noted in Sect. 3.2, the sky continuum
could indeed be two times brighter than assumed, based on a comparison
between Mauna Kea and Paranal sites. This would directly impact the
achieved SNR by a factor $\sqrt{2}$, e.g., a mean SNR of 10 would
translates into a mean SNR of 7.1. On the other hand, an emissivity of
15\% (pure Al coating) would imply a three times higher thermal
background, reducing the SNR even further, by a factor $\sqrt{3}$: a
mean SNR of 10 would translate into a mean SNR of 5.8. These different
assumptions directly impact the achieved SNR at a given redshift and
mass. For instance, this would lower the redshift up to which the GSMF
can be probed down to $M_{stellar}$, from z=4.1 down to
z$\sim$3.8-3.9. Therefore, while this could impact observations on a
galaxy-by-galaxy basis, it would not impact them strongly in a
statistical sense.

One of the current limitation of the simulations is that most of
inputs are currently considered as invariant with wavelength (e.g.,
throughput, emissivity), because the design of the telescope and the
instrument are not fully known at the moment. Simulations will need to
be updated once these characteristic curves will be known. However, we
expect this to be a second order effect, as characteristic curves are
known to vary only very specific small windows (e.g., absorption
features in optics transmission curves).

Finally, it is worth noting that we assumed a photon-starved sky
subtraction: sky and science frames were simulated using a different
noise realisation, which generated random errors only. This choice was
driven by the fact that, in the NIR, galaxies are usually selected in
redshift such that their emission lines fall in spectral windows free
of OH sky lines. In this case, the sky subtraction process can
reasonably well be assumed to be limited by the photon noise on the
sky continuum. This assumption implies that sky subtraction is not
affected by systematic errors. Further simulations related to the data
reduction strategy will be required to assess the impact of such
biases (affecting, e.g., measurements and/or estimation of sky
frames).

\subsubsection{Scientific inputs}
In reality, the morpho-kinematic range of templates is significantly
broader than the one used for the simulations. Because our knowledge
of galaxy formation is still incomplete, the limited number of
templates was chosen in order to cover a wide range of surface
brightnesses. We have intentionally focused on a few representative
cases, ranging from major mergers with complex morpho-kinematics to
irregular galaxies with flat velocity fields, and regular rotating
disks with different velocity gradients. Despite their limited number,
these templates should span most interested cases.

The impact of the different surface brightness distributions can be
seen in the middle panel of Fig. \ref{masslim}, where error-bars
reflect the range of mass that can be reached at a given redshift and
S/N for the 10 different morpho-kinematic templates considered in this
work. We find a tendency for the most irregular surface brightness
distributions to result in a largest mass limit at a given total flux.
Such surface brightness distributions result in a more complex
coupling between adjacent spectra in the IFS at limited spatial
resolution, which in turn results in emission lines with a lower peak
and larger wings compared to more regular distributions in surface
brightness.

Results from high resolution imaging and spatially-resolved
spectroscopy have revealed that high-z galaxies have very different
properties in their morphology and kinematics, compared to local
galaxies (e.g., \citealt{elmegreen05,forsterschreiber09}). In
particular, these observations suggest that a significant fraction of
high-z galaxies would be rotating clumpy disks (e.g.,
\citealt{genzel08}). We find that at z$\sim$2, the mass limit of
clumpy disks is roughly one order of magnitude larger than more
regular galaxies. Transposing this result on current 3D samples
observed with 8-10m telescopes, this could explain why very distant
galaxies detected by 3D spectroscopy so far, which are largely
dominated by irregular surface brightness distributions (e.g.,
\citealt{forsterschreiber09}), systematically show very large
equivalent widths in their emission lines to be detected. If a
significant fraction high-z galaxies are indeed clumpy disks, then the
true average SNR at a given redshift and mass might be biases toward
the largest values in the range shown in Fig. \ref{masslim}.

Most of physical quantities used as inputs are not known to a factor
better than two. We first discuss the source of uncertainty associated
with the intrinsic scatter of the scaling relations used to rescale
the morpho-kinematic templates in terms of flux and size. The
empirical relations between stellar mass and K-band magnitude as a
function of redshift have typical uncertainties of 0.3 dex (i.e., 0.75
mag, but likely more at z$\geq$4, see Sect. 3.1). This uncertainty on
the stellar mass directly translates into a 0.1 dex uncertainty on
$\log{R_{half}}$, using the scaling relation between stellar mass and
half-light radius $R_{half}$ adopted in Sect. 3.1. This is to be
compared to the intrinsic scatter of this relation (0.3 dex in
$\log{R_{half}}$, see \citealt{courteau07}), to the scatter of the
size versus wavelength relation (0.04 dex in $\log{R_{half}}$, see
\citealt{barden05}), and to that of the size versus redshift relation
($\sim$0.06 dex in $\log{R_{half}}$, see
\citealt{bouwens04,ferguson04,dahlen07}). The two dominant sources of
uncertainty are clearly the redshift vs. stellar mass / K-band
magnitude relation and the scatter of the stellar-mass vs. size
relation, with roughly an uncertainty of a factor two in size and
total flux.

Another source of uncertainty is the possible evolution of these
relations with redshift. While the stellar mass vs. size relation does
not appear to evolve, at least up to z$\sim$1 \citep{barden05}, the
Tully-Fisher relation appears to evolve smoothly with redshift (a
factor two in mass between z=0 and z$\sim$0.6, see \citealt{puech08},
slightly more at z$\sim$2, see \citealt{cresci09}). In this paper, we
assumed no evolution of this relation: this is conservative as, at a
given stellar mass, high-z rotating disks would have larger rotation
velocities, which would make the associated velocity gradients easier
to be detected. The scatter associated to the stellar mass
Tully-Fisher Relation is $\sim$0.15 dex in $\log{M_{stellar}}$,
meaning that it will not introduce significant additional uncertainty
on the stellar mass / K-band magnitude compared to the stellar mass
vs. size relation (see above).

Finally, one of the less constrained physical parameter is perhaps the
emission line equivalent width in very distant galaxies. This
parameter does not influence the flux distribution but only its total
intensity. We have adopted the median value found for [OII] at
z$\sim$1 (see Sect. 3.1 and 5.1.1), with an associated scatter of
15-20\AA, depending on sample selection \citep{hammer97}. Larger
values were observed at larger redshift, especially with H$\alpha$
\citep{erb06}. It is indeed expected that the typical equivalent
widths increase with redshift, as a result of the increase of the star
formation density with redshift, at least up to z$\sim$2
\citep{hopkins04}. In this paper, we chose not to parametrise emission
line equivalent width as a function of redshift. In practice, such a
parametrisation would be largely affected by the sample selection
criteria at very high-z, and the physical processes driving and
affecting emission within different lines, as one has to switch from
H$\alpha$ to [OII] above z$\sim$4. We therefore decided not to
introduce any additional uncertain parameter and adopted the observed
[OII] median value at z=1 as a conservative value. Moreover, adopting
a constant value helps to disentangle the different sources of surface
brightness evolution with redshift, which are only due to mass and
size evolution in this study. The results given in Sect. 4.2 are to
this respect conservative, especially between z$\sim$2 and 4, where
observations are not strongly limited by the thermal emission from the
telescope: considering an equivalent width of 100\AA~ in H$\alpha$, as
it is often observed in z$\sim$2 galaxies (e.g.,
\citealt{forsterschreiber09}), instead of 30\AA, would imply an
increase by a factor 3 of the mean SNR (see Sect. 4.2).

We therefore conclude that simulations are internally-consistent
within a factor two in all the physical quantities considered. For
instance, the empirical relations between stellar mass and K-band
magnitude are uncertain within a factor two, but the flux correction
between the rest-frame wavelength corresponding to the K-band and that
of the emission line considered is also less than a factor two (see
Sect. 3.1).

Finally, in this DRM program, we assumed that targets will be selected
from future deep and complete optical-NIR imaging and redshift
surveys. Such surveys could potentially be done using existing
facilities such as VISTA or the LSST, but will probably also benefit
from the follow-up of future instruments such as JWST/NIRSPEC or ALMA.
To guarantee the representativity of the sample and optimise the
sampling of the look-back time axis, we did not assume any
pre-selection in colour, nor do we pre-select redshifts in terms of
atmospheric transmission as it is currently done on 8-10m telescopes.
This allows us to take advantage of the huge collecting power of the
E-ELT, and target any potential galaxy regardless of its redshift. As
an alternative observational strategy, one could select more optimal
redshift windows in correspondence of atmospheric windows, which would
directly result in a gain of SNR and survey speed. Looking at Fig. 6,
optimal windows are around 1.6 $\mu$m (which corresponds to z$\sim$3.3
using [OII]), and 2.2 $\mu$m (which corresponds to z$\sim$2.3 using
H$\alpha$ or 4.9 using [OII]. Selecting galaxies in these windows
would guarantee an atmospheric transmission $\sim$95\% (provided that
targets are selected between strong absorption features), i.e.,
roughly a factor two larger than the atmospheric transmission
corresponding to simulations at z=2, with 44\%, and at z=4, with 63\%:
for instance, targeting galaxies at z$\sim$2.3 instead of z=2, would
result in a gain of 2.2 in SNR. Such a factor is well below the
inherent uncertainty associated to the simulations (see above).

\subsection{Impact of telescope design and size}

The telescope is by far the dominant source of background in the
K-band. Because the SNR is in a background-limited regime at z=5.6
(see Fig. \ref{masslim}), the telescope emission limits source
detection at very high redshift. Note that in the simulations, we have
used an optimistic assumption for the emissivity of the 5-mirror E-ELT
design, with $\epsilon _{tel}=$5\%\footnote{see
http://www.eso.org/sci/facilities/eelt/science/drm/tech\_data/telescope/}.

MOAO provides only partial seeing corrections, therefore the resulting
PSF is dominated by residual atmospheric perturbations
\citep{assemat07}. This implies that the telescope diameter does not
directly influence the spatial resolution of observations, but only
the achieved SNR. The scaling relation of Sect. 4.2 shows that there
is no breaking point in telescope diameter: with a smaller 30m
telescope, one would need two times longer exposures to reach the same
SNR.

In terms of survey speed, the E-ELT represents a huge improvement
compared to existing 3D spectrograph on 8-10m telescopes. The fiducial
program depicted in Sect. 4.3 will deliver 75 galaxies at z$\sim$2
covering the whole mass spectrum in only 2.3 nights. It is worth
comparing this number with, e.g., the SINS survey at the VLT.
\cite{forsterschreiber09} described this survey of 63 galaxies and
gave an average integration time of 3.4 hr per galaxy (per band),
which represent about 27 VLT nights in total. Note that this does not
account for observing time lost because of undetected targets. The
E-ELT will allow us to observe roughly the same number of galaxies,
but saving more than a factor 10 in time. In addition, one should
consider that SINS galaxies have typically an emission line equivalent
width of $\sim$100\AA, while the simulations assume 30\AA. Rescaling
simulations with a gain of 3 in equivalent width leads to a gain of 9
in integration time. It basically means that the SINS survey could be
completed in less that one night with the E-ELT. Moreover, the E-ELT
will enable a better control of the selection function: because of the
limited sensitivity of current 8-10m telescopes, pre-selection in
colour must be used to draw 3D distant samples beyond z$\sim$1 (see
Sect. 2.1). In comparison, the E-ELT will allow us to conduct a very
complete 3D follow-up of a pure mass-selected sample. Sect. 4.2.1
demonstrates that the E-ELT will allow us to draw such samples up to
z$\sim$4-5. Such a limit will clearly remain out of reach of current
8-10m telescopes, except for a handful of targets.

\subsection{Impact of site properties}

The sky is the dominant source of background only for $z \leq 4$
observations. Note that in the simulations we have used a sky model
from Mauna Kea, which is $\sim$2 times fainter in the H band that the
official DRM Paranal model (see Sect. 3.2 and Fig. \ref{atm}). As $SNR
\propto 1/\sqrt{background}$ in such a regime, a site having two
(four) times higher background than the one used in the simulations
will reduce the achieved SNR from, e.g., 10 to 7 (5). The impact on
high-redshift observations is therefore relatively limited (see Fig.
\ref{masslim}).

The SNR loss between observations with a seeing of 0.8 arcsec and
observations with a seeing of 0.95 arcsec is 5-15\%. The strongest
impact of worst seeing conditions will be to limit the capability to
recover Rotation Curves and detailed kinematics of distant galaxies.

\subsection{Impact of Instrument}
GLAO limits the achievable SNR to lower values, compared with MOAO. As
a consequence, the recovery of the large scale motions in distant
galaxies using GLAO is limited to higher masses. Any further objective
of the DRM will be strongly limited in using GLAO instead of MOAO
(e.g., rotation curves, detailed kinematics).

Appropriate targets in the NIR are usually selected such as they have
emission lines that fall in regions free of strong OH lines. This
requires a minimal spectral resolution of R$\sim$3000 to resolve OH
sky lines with enough accuracy. On the other hand, the highest the
spectral resolution, the better the accuracy on the recovered
kinematics (see Tab. \ref{tabacc}): at least R=5000 is required if one
wants to recover the velocity dispersion with no more that 50\% of
relative uncertainty. This value appears to be a good compromise
between the desire to minimise the impact of the OH sky lines and not
wanting to over-resolve the line by a large factor: the scaling
relation between SNR and R (see Sect. 4.2) demonstrates the interest
of having the smallest spectral resolution possible, which optimises
the achieved SNR.

The choice of the IFU pixel scale drives the spatial resolution of
MOAO-fed 3D spectroscopy observations, as in the range of EE provided
by MOAO, the PSFs have a FWHM largely smaller than twice the pixel
size. Therefore, the choice of the optimal IFU pixel scale is related
to the optimal ``scale-coupling'' between the IFU pixel scale and the
spatial scale of the physical feature that one wants to recover using
this IFU \citep{puech08b}. This can be quantified by using the ratio
between the size of this feature (here, the galaxy diameter, as one
wants to recover the large-scale rotation) and the size of the IFU
resolution element. 3D observations of z$\sim$0.6 galaxies with
FLAMES/GIRAFFE have demonstrated that a scale-coupling of at least 3
can already provide us with useful information on large-scale motions
\citep{flores06}, allowing us to distinguish rotating disks from other
more complex systems \citep{puech06,yang08}. \cite{epinat09b} showed
that such a scale-coupling is enough to provide us with unbiased
estimated of the PA and rotation velocity in distant galaxies. It
corresponds to the minimum value necessary to ensure that each side of
the galaxy is at least spatially sampled by the IFU at the Nyquist
rate. Of course, there are specific situations where a finer spatial
sampling is desirable in order to properly distinguish between complex
systems and rotating disks (see, e.g., \citealt{epinat09b}), but using
this as a guide-line provide us with a useful upper limit on the IFU
spatial sampling, which is optimised for very high surface brightness
detection. In the z=4 simulations, this minimal scale-coupling leads
to a pixel scale of ~55mas, 125 mas, and 280 mas for 0.1, 1, and 10 M*
galaxies, respectively. Hence, a minimal pixel scale of 50 mas is
required if one wants to be able to recover, at least in principle,
large-scales motions in z=4 galaxies, and for a large range of
stellar-mass.

Figure \ref{sim} shows the velocity fields, velocity dispersion maps,
and emission line maps extracted from the simulations using a 50 mas
pixel scale. It is clear that the lowest mass case barely provides
enough spatial information to clearly distinguish between the rotating
disk and the major merger. Even if the scale-coupling is in principle
large enough to properly recover large-scale motions, the pixel scale
limits the achieved SNR to relatively low values (see
Tab. \ref{tabsnr}). If one only wants to recover large-scale motions
in the most massive objects, then an IFU pixel scale of 75 mas can be
used, providing more SNR at constant integration time (see
\citealt{puech08b}).

Finally, we adopted 50 mas as a reference pixel scale in the
simulations. This resulted from a compromise between resolving sub-kpc
scales (100 mas represent 0.7 kpc at z=4) and the resulting surface
brightness sensitivity. We derived the emission line flux sensitivity
per pixel in the following way. We considered only simulations with 50
mas per pixel and R=5000 (the result can be easily rescaled to any
other couple of spatial and spectral sampling). Then, we constructed
the histogram of the emission line flux corresponding to a threshold
of $SNR_{kin}$ larger than three (which defines the spaxels used for
analysing the VF, see Sect. 2.1). In this histogram, we considered all
simulations with integration times of 24 hr, but with different
seeing, AO correction and morpho-kinematic templates, which allow us
to average over a large number of different observational conditions.
We found that this histogram peaks at 2$\times 10^{-20}$erg/s/cm$^2$.

In the simulations, we also explored a 25 mas pixel scale, which
allowed us to bracket the EAGLE baseline with 37.5 mas. Compared to
our reference case, the EAGLE baseline (R=4000 and 37.5mas/pix) would
imply a 16\% decrease in SNR per spatial and spectral element of
sampling (see Sect. 4.2.1), and degrade the emission line flux
sensitivity limit by a factor 2.2. However, the EAGLE pixel scale
provides an optimal scale-coupling with distant clumps that are
ubiquitous in distant galaxies \citep{elmegreen05}: such clumps are
typically 1 kpc is diameter, therefore such a pixel scale would
provide us with a scale-coupling of 3-4 between z=2 and z=4, which is
an optimal compromise for studying the structure and kinematics of
these clumps \citep{puech09b}. The SNR per pixel goes linearly with
the pixel scale: adopting a 4 mas pixel scale, which corresponds to
the diffraction limit in H band, would imply a loss of 92\% in SNR per
pixel compared the 50 mas pixel scale. This has some bearing on
detailed studies of relatively nearby galaxies (and in particular of
their central AGN) for which the E-ELT will provide a huge improvement
compared to, e.g., the VLT (see Sect. 5.2), as it will allow us to
probe finer spatial scales (8 mas represents 0.07 kpc at z=2).
However, such a finner spatial sampling is much less efficient in
terms of survey speed for observing higher redshift galaxies.
Moreover, the structure of distant galaxies at such small spatial
scales remains uncertain.

\begin{table}
\centering
\begin{tabular}{cccccc}\hline
SNR & 3-4 & 4-5 & 5-7 & 7-10 & $\geq$10\\\hline
R=2500  & 70km/s & 46km/s & 31km/s & 25km/s & 9km/s-\\
        & 73\% & 59\% & 42\% & 21\% & 9\%\\
R=5000  & 59km/s & 38km/s & 31km/s & 21km/s & 8km/s\\
        & 50\% & 42\% & 29\% & 20\% & 8\%\\
R=10000 & 54km/s & 37km/s & 57km/s & 20km/s & 8km/s\\
        & 50\% & 36\% & 26\% & 18\% & 7\%\\\hline
\end{tabular}
\caption{Accuracy on the velocity (in km/s, first lines) and relative
  accuracy (in \%, second lines) on the velocity dispersion
  measurement as a function of spectral resolution (Monte-Carlo
  simulations).}
\label{tabacc}
\end{table}

\section{Conclusions}
We have conducted simulations of the ``Physics and mass assembly of
galaxies out of z$\sim$6'' science case for the E-ELT, exploring a
wide range of observational and physical parameters. We have defined
figures of merit for this science case despite the inherent complexity
of the science goals, derived empirical scaling relations between the
signal-to-noise ratio of kinematic and intensity maps and the main
telescope and instrument parameters, as well as a relation between the
limit in stellar mass that can be reached for a given signal-to-noise
ratio as a function of redshift. We specifically investigated the
impact of AO performance on the science goal. We did not identify any
breaking points with respect to all parameters (e.g., the telescope
diameter), with the exception of the telescope thermal background,
which strongly limits the performance in the highest (z$>$5) redshift
bin. We find that the full range of science goals can be achieved with
a $\sim$100 nights program on the E-ELT, provided a high multiplex
advantage $M\sim N_{gal}/8$. We stress that several assumptions and
guided guesses had to be made on both the observational conditions and
physical characteristics of distant galaxies under study. This
introduces an inherent uncertainty, which can be mitigated with future
simulations as the telescope design will be consolidated and more
details about the physics of high-z galaxies will become available.

\section*{Acknowledgements}
We are especially indebted to T.J. Cox and F. Bournaud who provided us
with the hydro-dynamical simulations of merging galaxies and clumpy
disks respectively, to P. Amram and B. Epinat who provided us with
kinematic data of local galaxies from the GHASP survey, as well as to
I. Fuentes-Carrera, who has provided us with Fabry-Perot data of
ARP271. We thank N. F\"orster-Schreiber who kindly provided us with
the reduced data-cube of their SINFONI observations of BzK-15504. M.P.
wishes to thank R. Gilmozzi for financial support at ESO-Garching,
where this work was carried out. S.T. greatfully acknowledges support
from the Lundbeck foundation. We thank M. Franx, I. Hook, J. Bergeron
and all the ESO E-ELT Science Working Group, as well as the E-ELT
Project Office, the E-ELT Science Office (EScO), and J.-G. Cuby for
useful discussions regarding the subject of this paper. We especially
thank J. Liske for very useful discussions concerning the modelling of
the sky background. This work received the support of PHASE, the high
angular resolution partnership between ONERA, Observatoire de Paris,
CNRS and University Denis Diderot Paris 7, as well as of the ``Agence
Nationale de la Recherche'' (ANR-06-BLAN-0191).

\label{lastpage}

\end{document}